\documentclass[useAMS,usenatbib]{mn2e}
\usepackage{amsmath}
\usepackage{url}
\usepackage{graphicx}

\def\dim#1{\mbox{\,#1}}

\font\japit = cmti10 at 10truept

\title
     [Intergalactic Filaments]
{\vglue-3.0truecm
\centerline{\japit Accepted for publication in Monthly Notices}
\vglue 2.5truecm
\noindent
Intergalactic Filaments as Isothermal Gas Cylinders
\author[A. G. Harford et al.]
	{A. Gayler Harford$^1$ and Andrew J. S. Hamilton$^{1,2}$ \\
	$^1$JILA, University of Colorado, Boulder, CO 80309, USA \\
	$^2$Dept.\ Astrophysical \& Planetary Sciences,
	University of Colorado,
	Box 440,
	Boulder CO 80309, USA}
}

\newcommand{\unit}[1]{\, {\rm #1}}

\newcommand{\bolder}[1]{#1}

\newcommand{\bold}[1]{#1}

\newcommand{\cmdscheme}{
    \begin{figure}
    \begin{center}
    \leavevmode
    \includegraphics[scale=0.29]{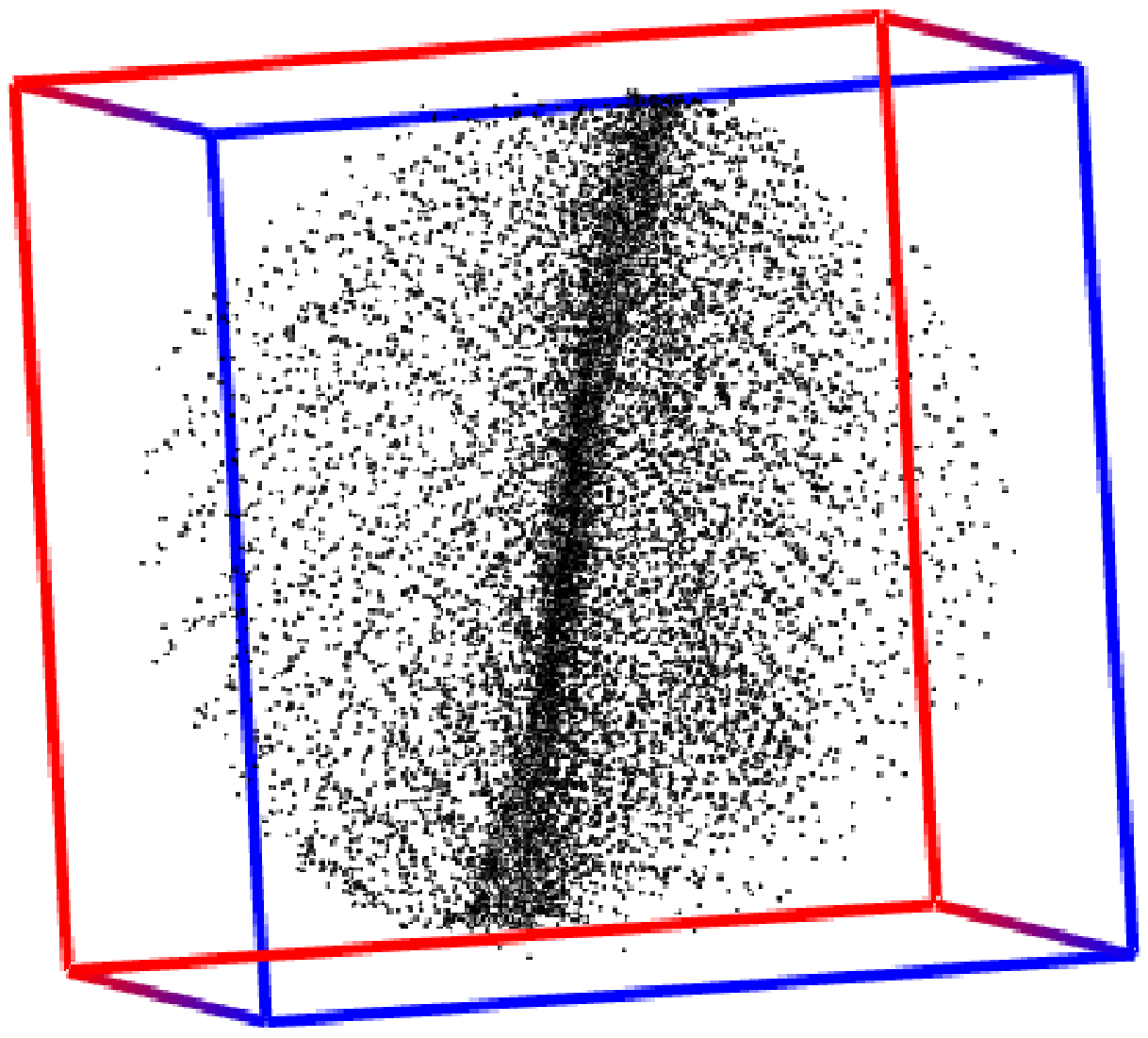}
    \includegraphics[scale=0.29]{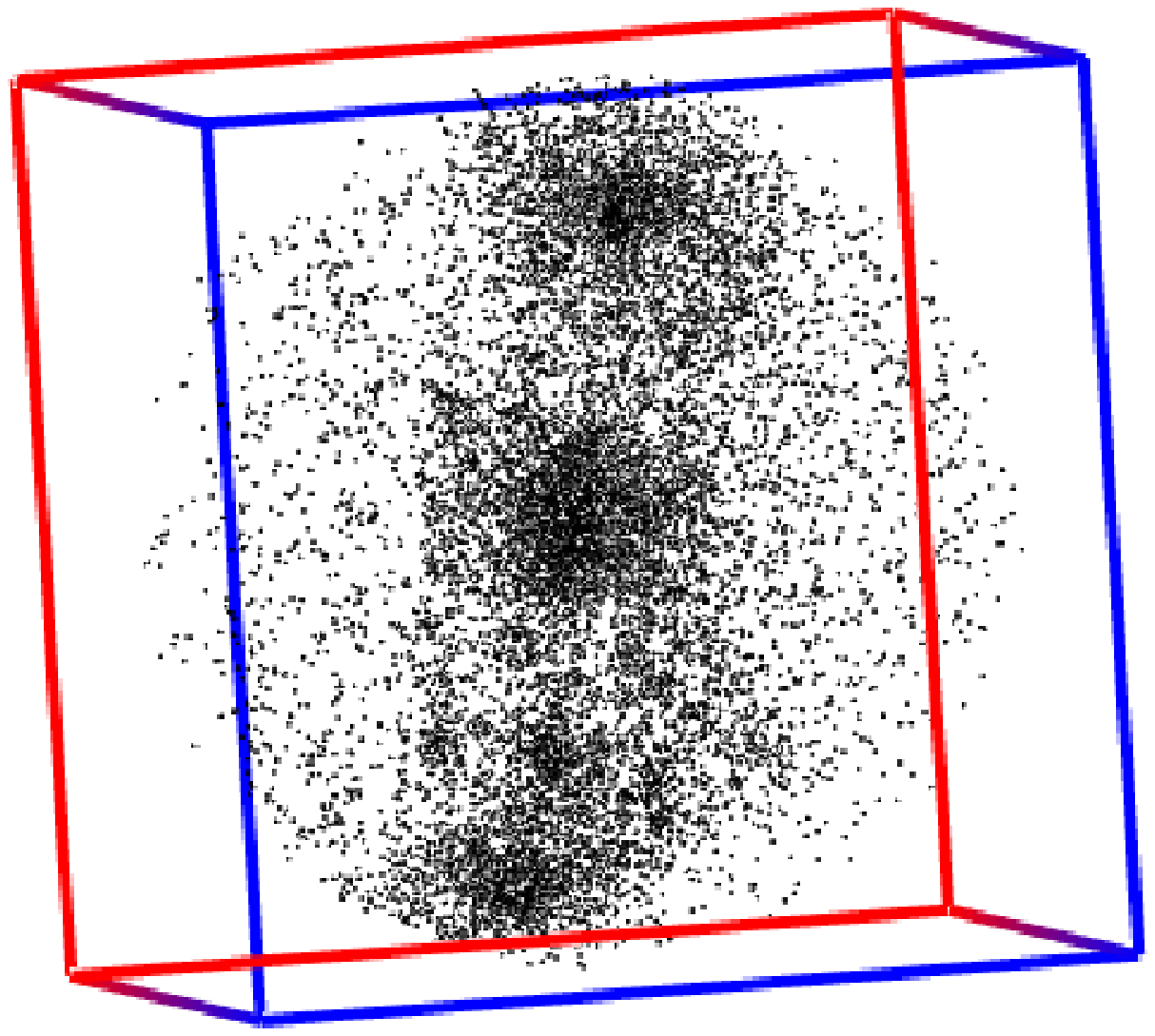}
    \includegraphics[scale=0.29]{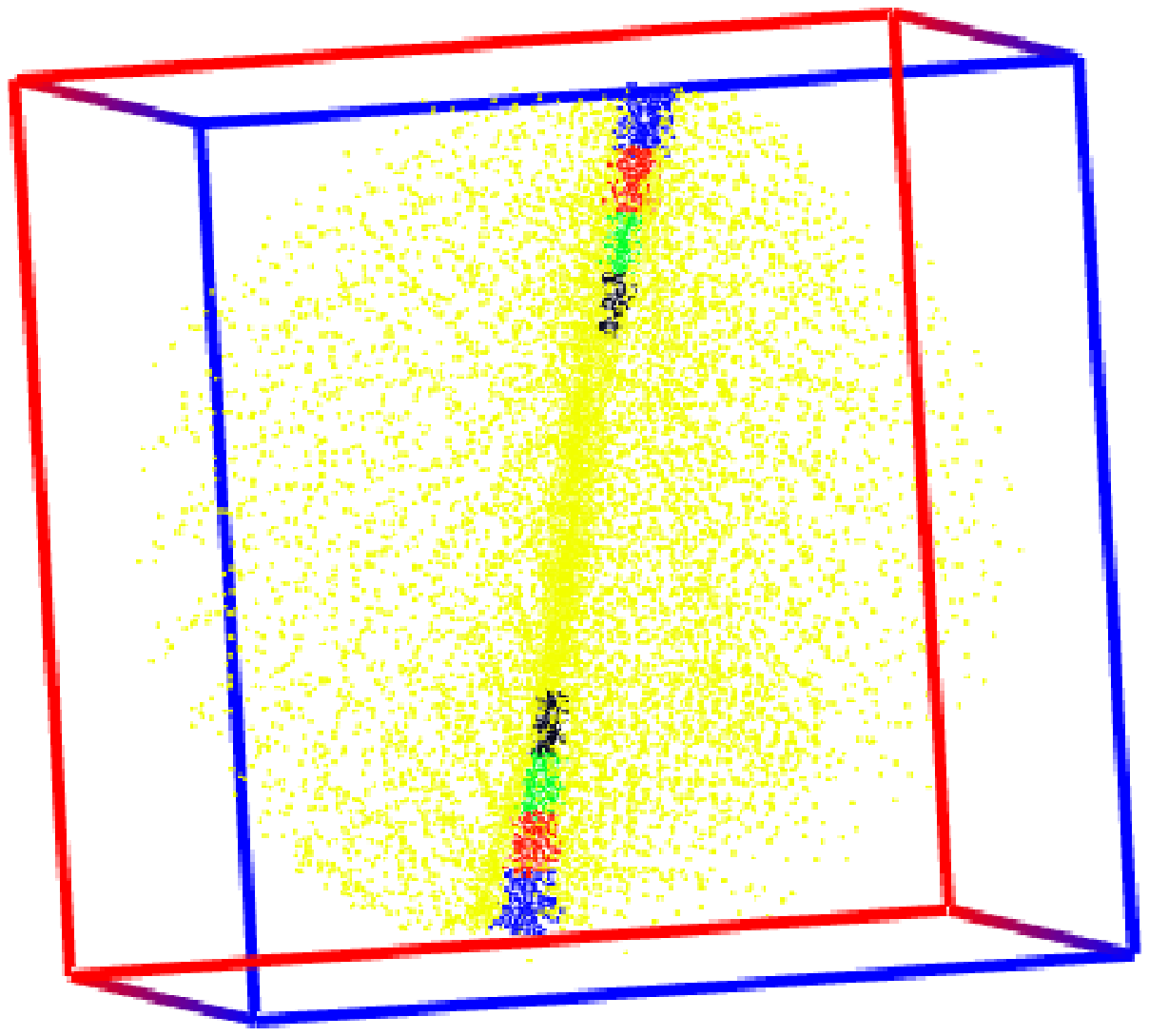}
    \includegraphics[scale=0.29]{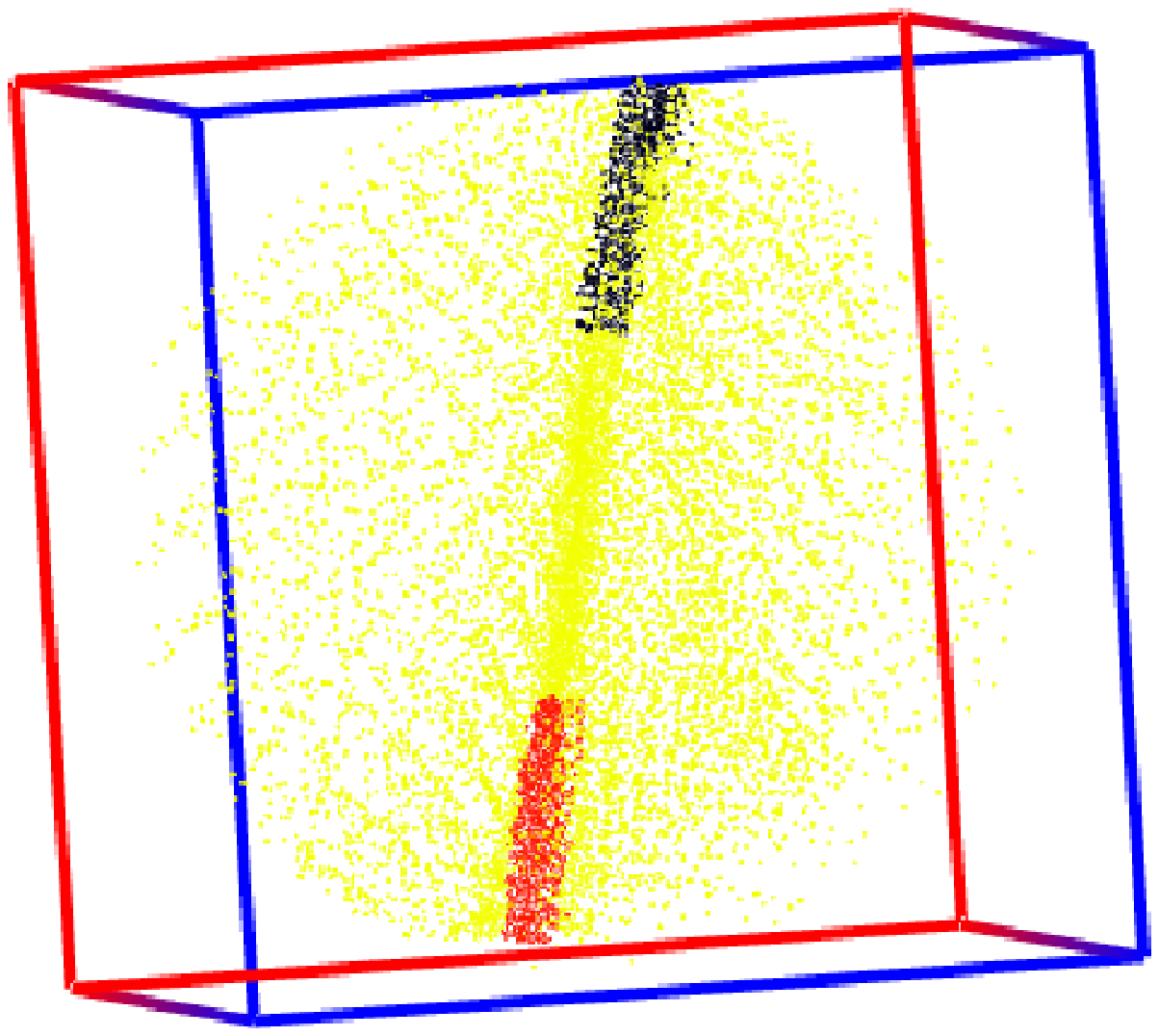}
    \end{center}
    \caption[1]{
    \label{scheme}
    \bold{Illustrative example of segment scheme.  Each image shows 
    simulation particles within a radius of about 51 kpc of the
    center of the same galaxy.  The images are enclosed in boxes
    to emphasize their three dimensional nature.  Top left: gas
    particles.  Top right:  dark matter particles.  
    Bottom left gas particles with each segment colored a
    different color.
    Bottom right:  gas particles with those from isothermal
    segments shown in black and those from non-isothermal
    segments shown in red.  The criteria for a segment to be
    isothermal are presented in Section~\ref{results} and
    in greater detail in Section~\ref{methods}}
    }
    \end{figure}
}

\newcommand{\cmdhighway}{
    \begin{figure}
    \begin{center}
    \leavevmode
    \includegraphics[scale=0.9]{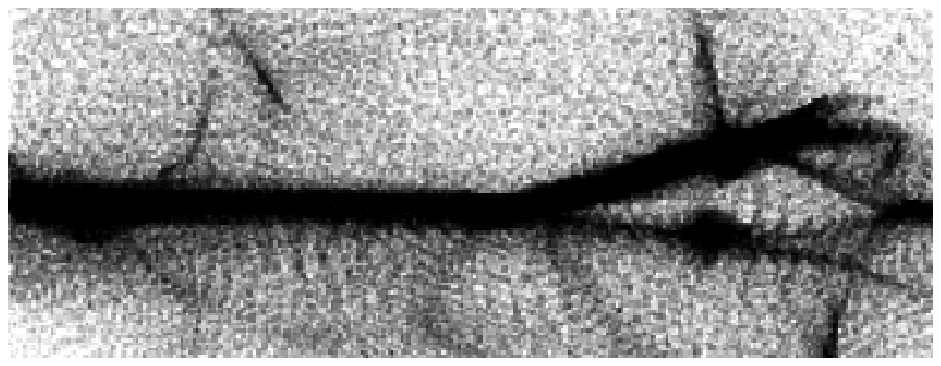}
    \includegraphics[scale=0.9]{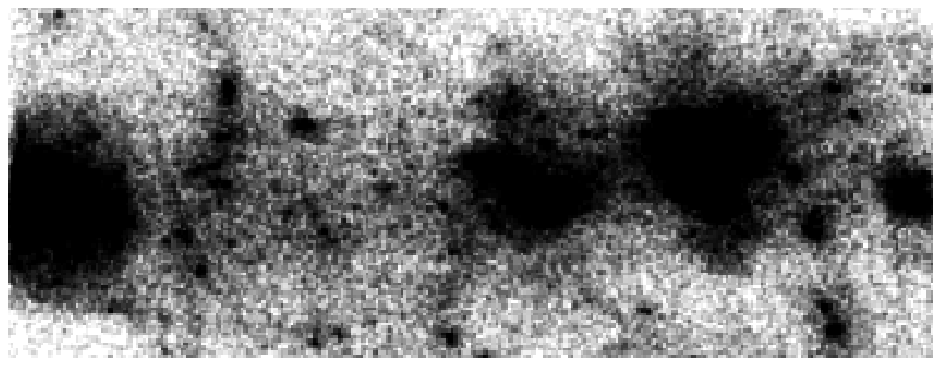}
    \end{center}
    \caption[1]{
    \label{highway}
    \bold{Filament in the simulation. 
    Shown is a filament about 240 kpc long.
    Top image shows just the gas particles.
    Bottom image shows just the dark matter particles.}
    }
    \end{figure}
}

\newcommand{\cmdgforig}{
    \begin{figure}
    \begin{center}
    \leavevmode
    \includegraphics[bb=50 0 504 360,scale=0.50]{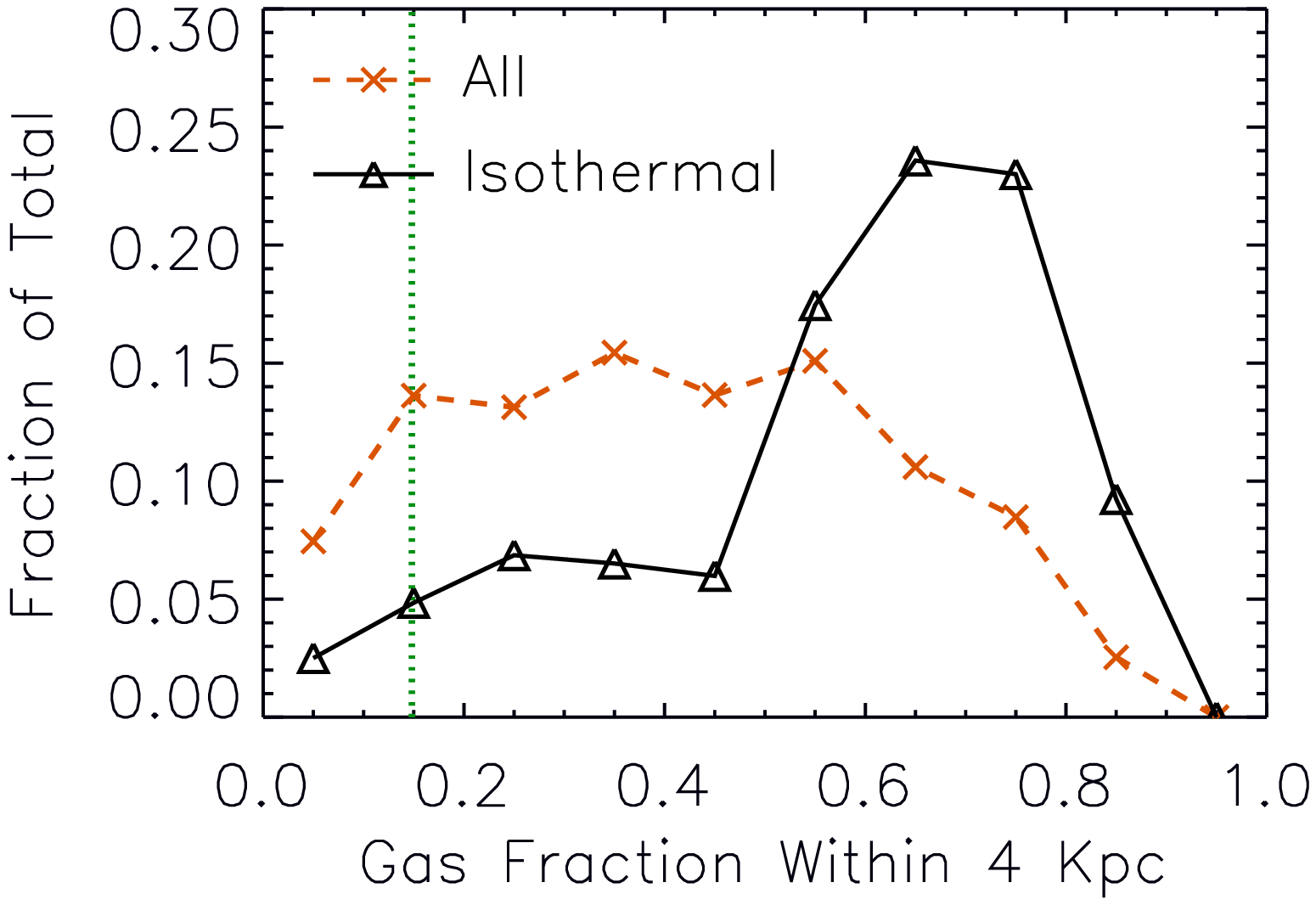}
    \end{center}
    \caption[1]{
    \label{gf_orig}
    \bold{Histogram of} gas fractions of segments.
    \bold{The bin size is 0.1.}
    Dashed, orange line with X's represents all
    \bold{segments studied.}
    Solid, black line with triangles represents
    the isothermal cylinders identified.
    The vertical, green, dotted line
    marks the mean cosmic baryon fraction.  The gas fraction is
    computed for all matter within 4 kpc of the axis of the segment.
    Stellar particles, which are treated as collisionless in the
    simulation, are combined with dark matter for
    this computation.
    }
    \end{figure}
}

\newcommand{\cmdcontours}{
    \begin{figure}
    \begin{center}
    \leavevmode
    \includegraphics[bb=50 0 504 360,scale=0.6]{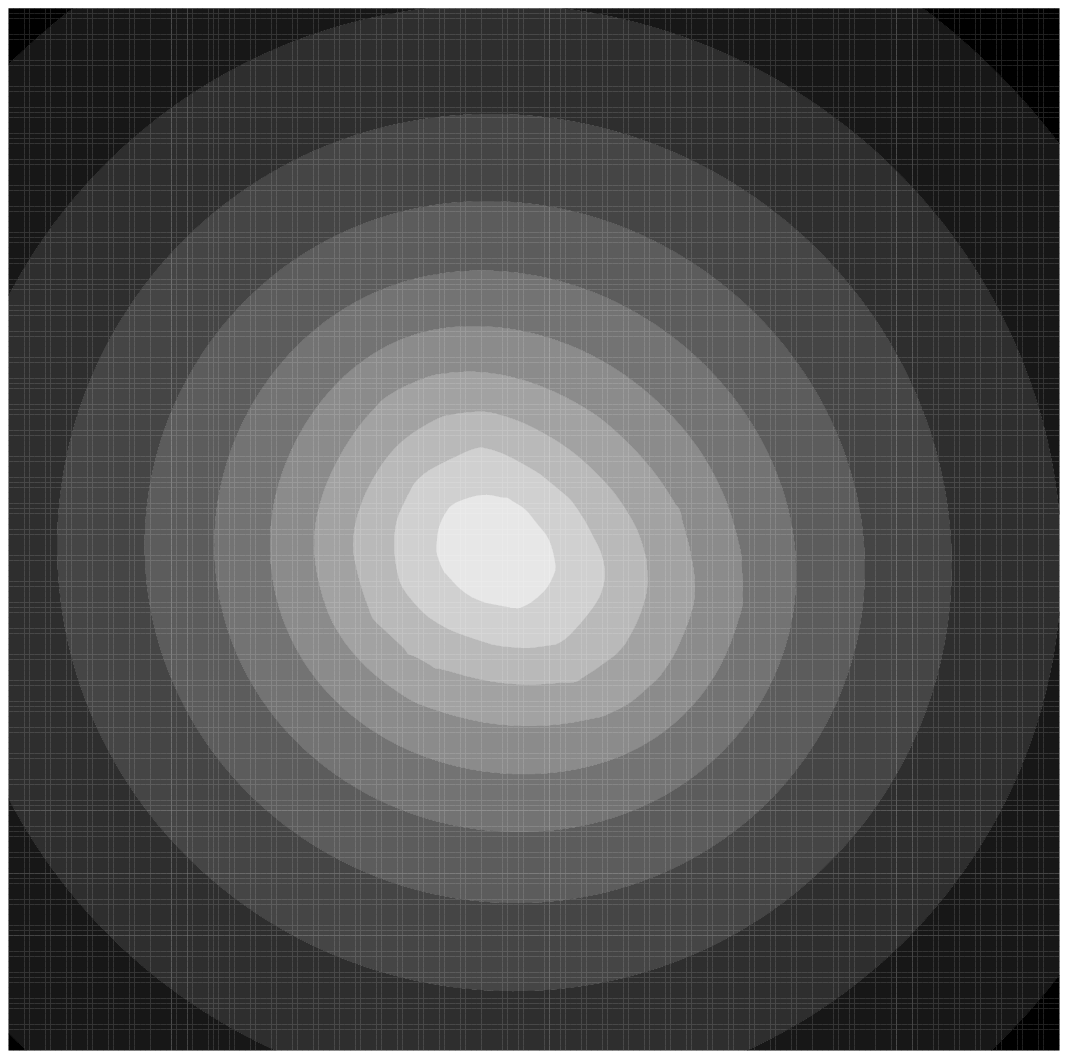}
    \end{center}
    \caption[1]{
    \label{contours}
    Gas potential contours for a segment of a filament of 
    the largest galaxy.
    The dimension of image is 34 kpc. 
    The potential is computed from
    gas within 7 kpc of central axis.  Fifty equally spaced contours
    are computed.  For clarity only a few of them 
    are shown here.
    }
    \end{figure}
}

\newcommand{\cmdrangefract}{
    \begin{figure}
    \begin{center}
    \leavevmode
    \includegraphics[bb=50 0 504 360,scale=0.55]{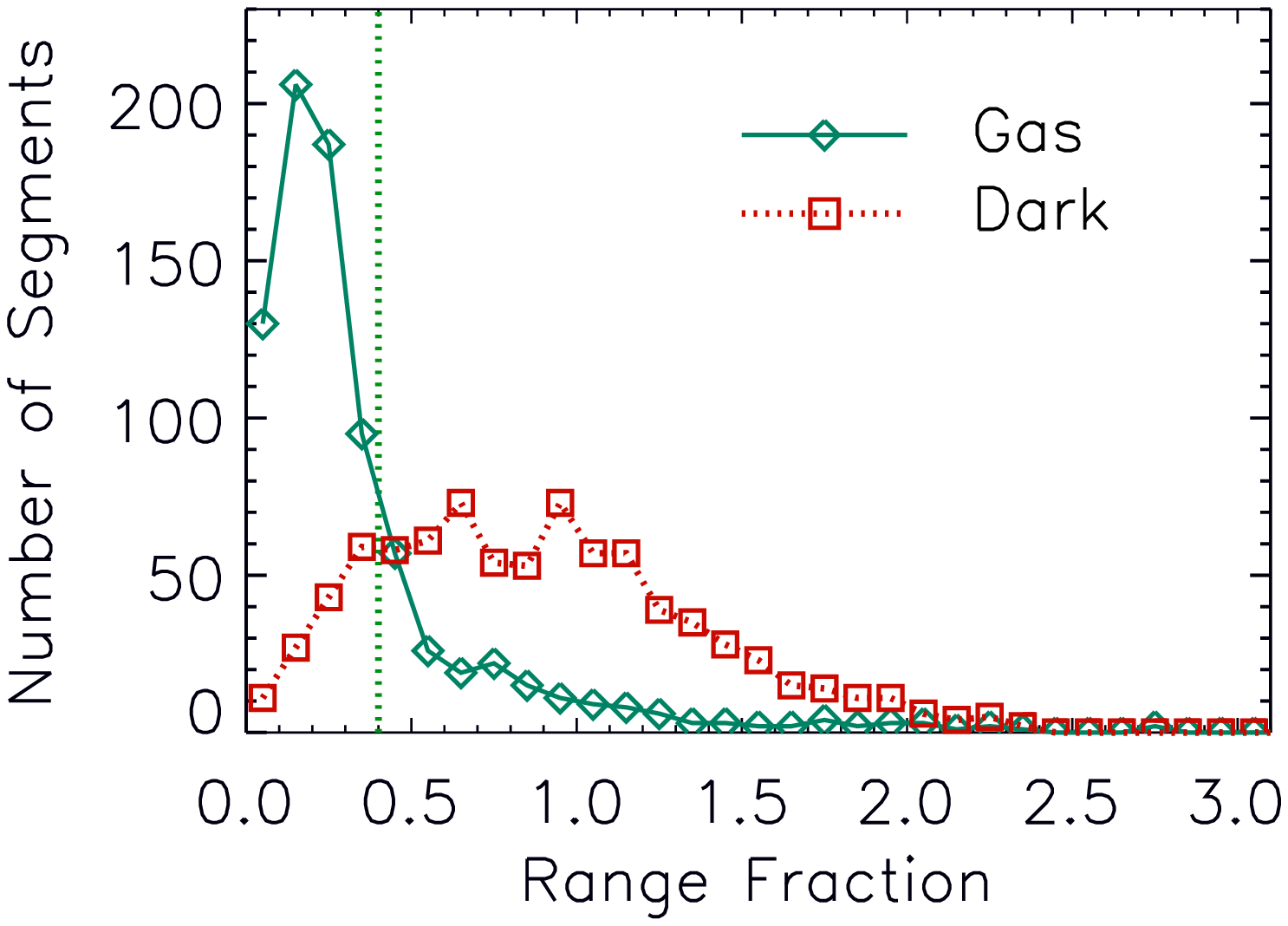}
    \end{center}
    \caption[1]{
    \label{range_fract}
    Histogram showing the distribution of gas and dark matter
    along segments.  \bold{The bin size is 0.125.}
    Solid, blue line with diamonds 
    represents the gas.  Dotted, red line with
    squares represents the dark matter.   
    After dividing the segment
    longitudinally into three equal parts,
    a ``range fraction'' is obtained by dividing the maximum
    difference in mass between any two parts by the 
    average mass of the three parts.  \bold{For this test we consider the
    mass within 4 kpc of the longitudinal axis.}
    We consider the distribution uniform if the range fraction does
    not exceed 0.4.  This cutoff is indicated by the green, dotted line.
    }
    \end{figure}
}

\newcommand{\cmdfit}{
    \begin{figure*}
    \begin{minipage}{175mm}
    \includegraphics[scale=0.3]{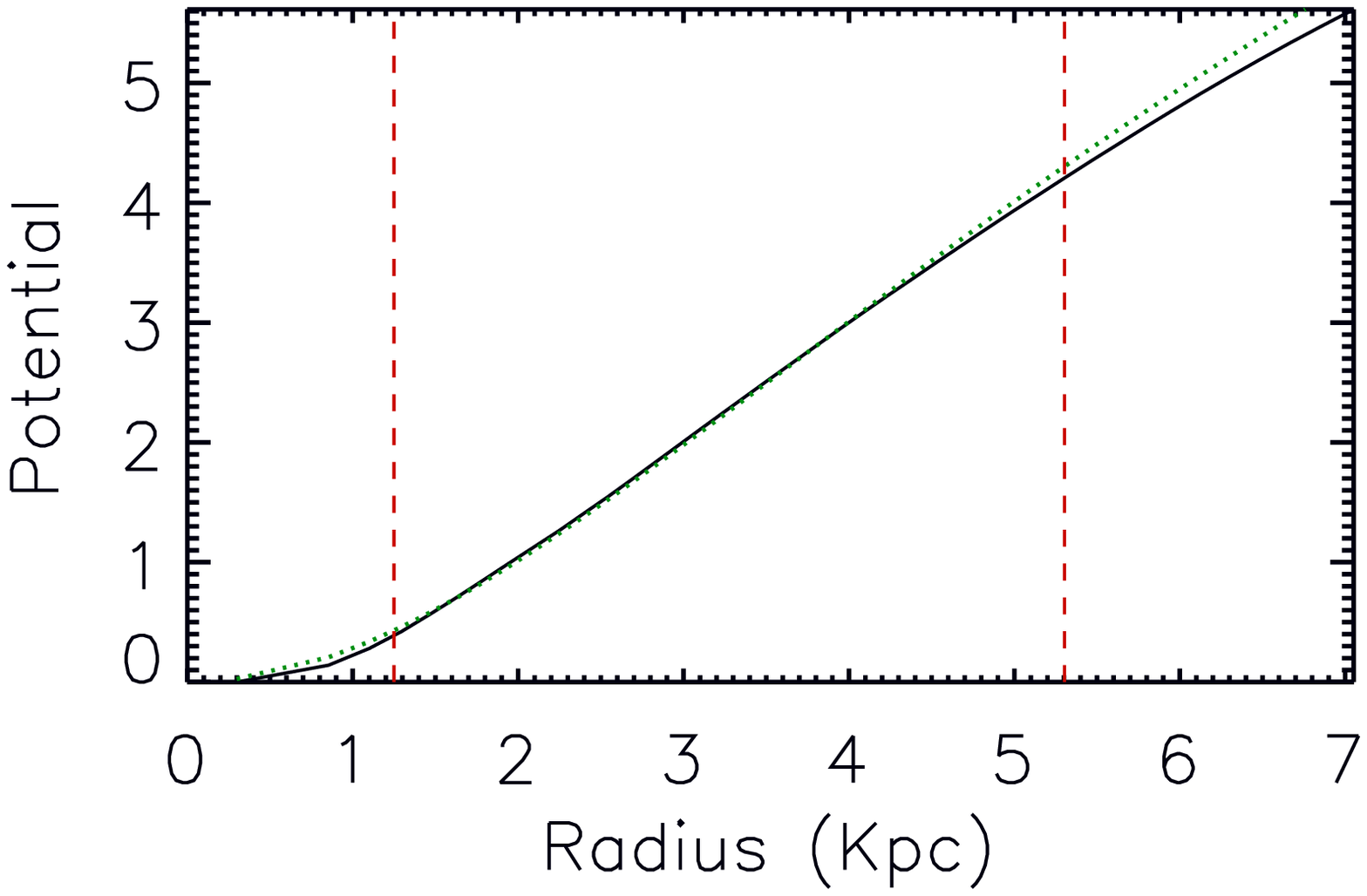}
    \includegraphics[scale=0.3]{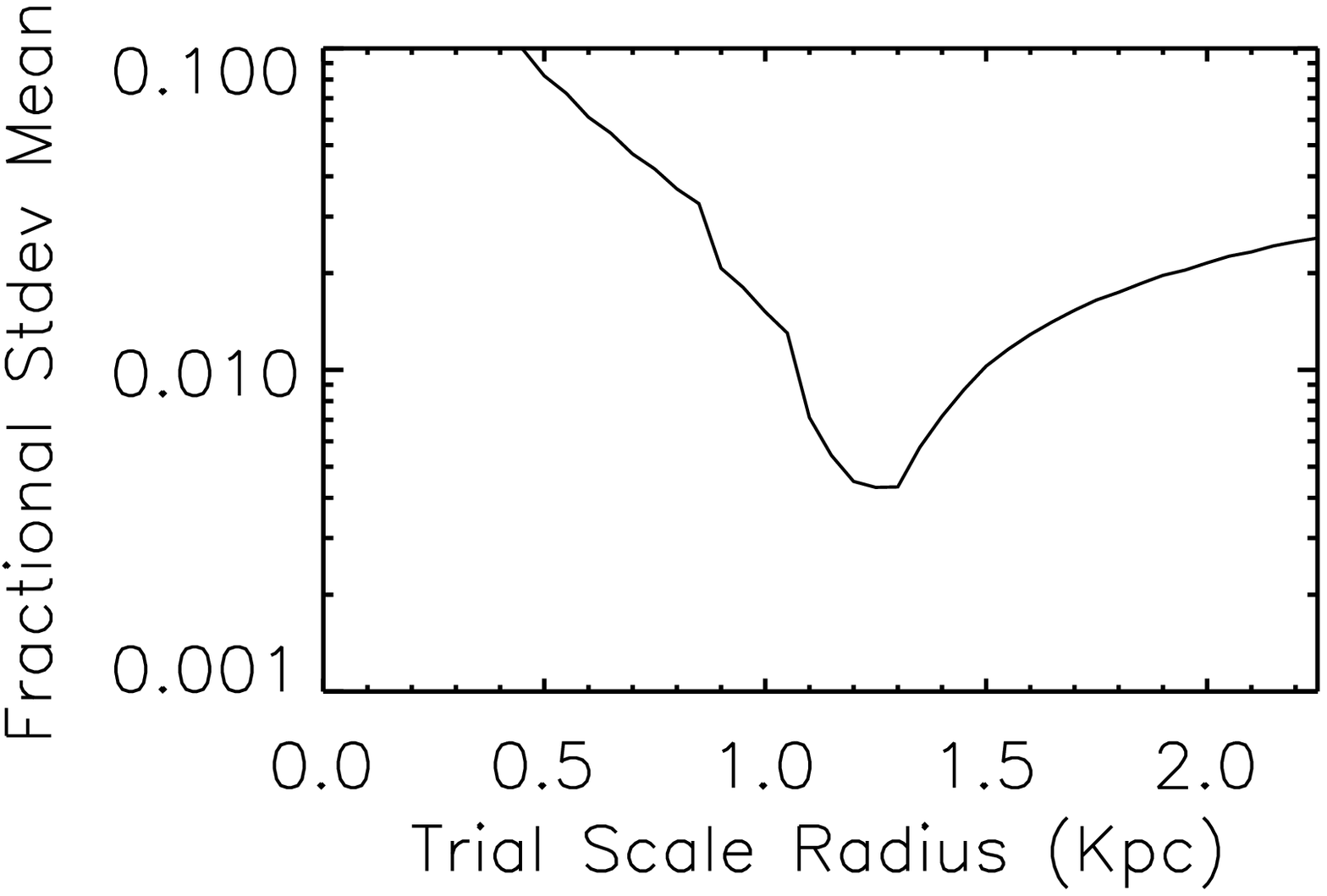}
    \includegraphics[scale=0.3]{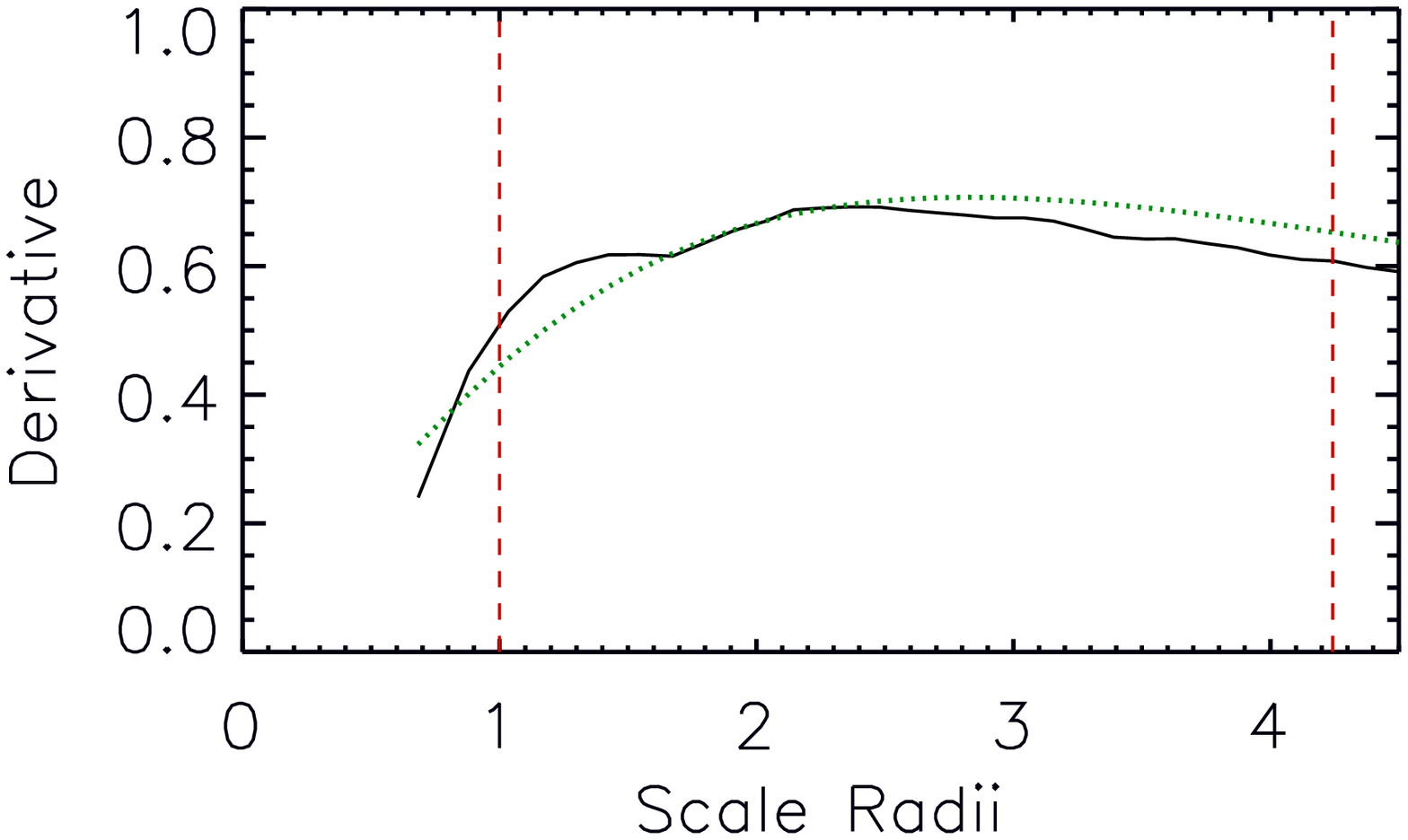}
    \caption[1]{
    \label{fit}
    Fitting data for a segment derived from
    a filament emanating from the largest galaxy in the simulation.
    In the left and right panels, the dotted, green lines are the 
    theoretical prediction.  
    The vertical, dashed, red
    lines enclose the fitting region for the optimal fit that
    is shown.

    Left:  Best fit of gas potential $\psi$ to theory.  
    Solid, black line 
    is the potential of the 
    segment divided by the square of the best-fit sound
    speed plotted as a function of radius.  See section~\ref{ration}
    for a definition of $\psi$.

    Center:  Delimitation of best fit.  Shows the ratio of the
    standard deviation of the mean of the proportionality constant
    to the constant itself as a function of the trial scale radius
    for the segment whose profile is shown on the left.

    Right:  Fit of the derivative of the gas potential $\psi$ 
    on the left to theory.  Solid, black line is the
    derivative of the gas potential shown in left panel.
    The abscissa shows the
    radius as a multiple of the scale radius for this fit.
    }
    \end{minipage}
    \end{figure*}
}
\newcommand{\cmdtmptr}{
    \begin{figure}
    \begin{center}
    \leavevmode
    \includegraphics[bb=50 0 504 360,scale=0.5]{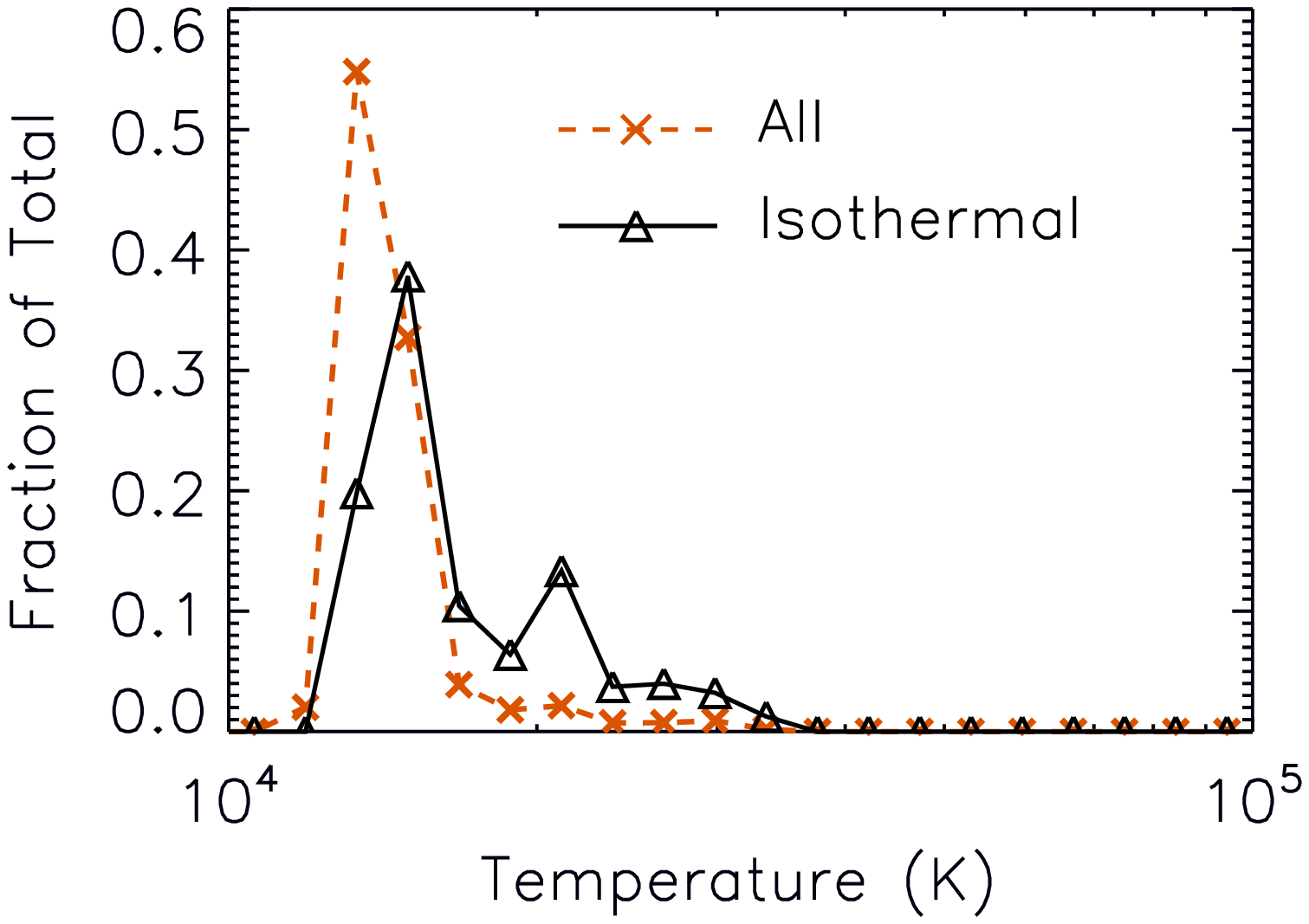}
    \end{center}
    \caption[1]{
    \label{tmptr}
    Histogram of average gas temperatures of segments.  \bold{The
    bin size is 0.05 on a logarithmic scale.}
    Dashed, orange line with X's represents all
    segments.  Solid line with triangles
    represents the isothermal segments identified
    in this study.
    The temperature was averaged by mass for the gas within 4 kpc of
    the axis of the \bold{segment.}
    }
    \end{figure}
}

\newcommand{\cmdshellssqrat}{
    \begin{figure}
    \begin{center}
    \leavevmode
    \includegraphics[bb=50 0 504 360,scale=0.55]{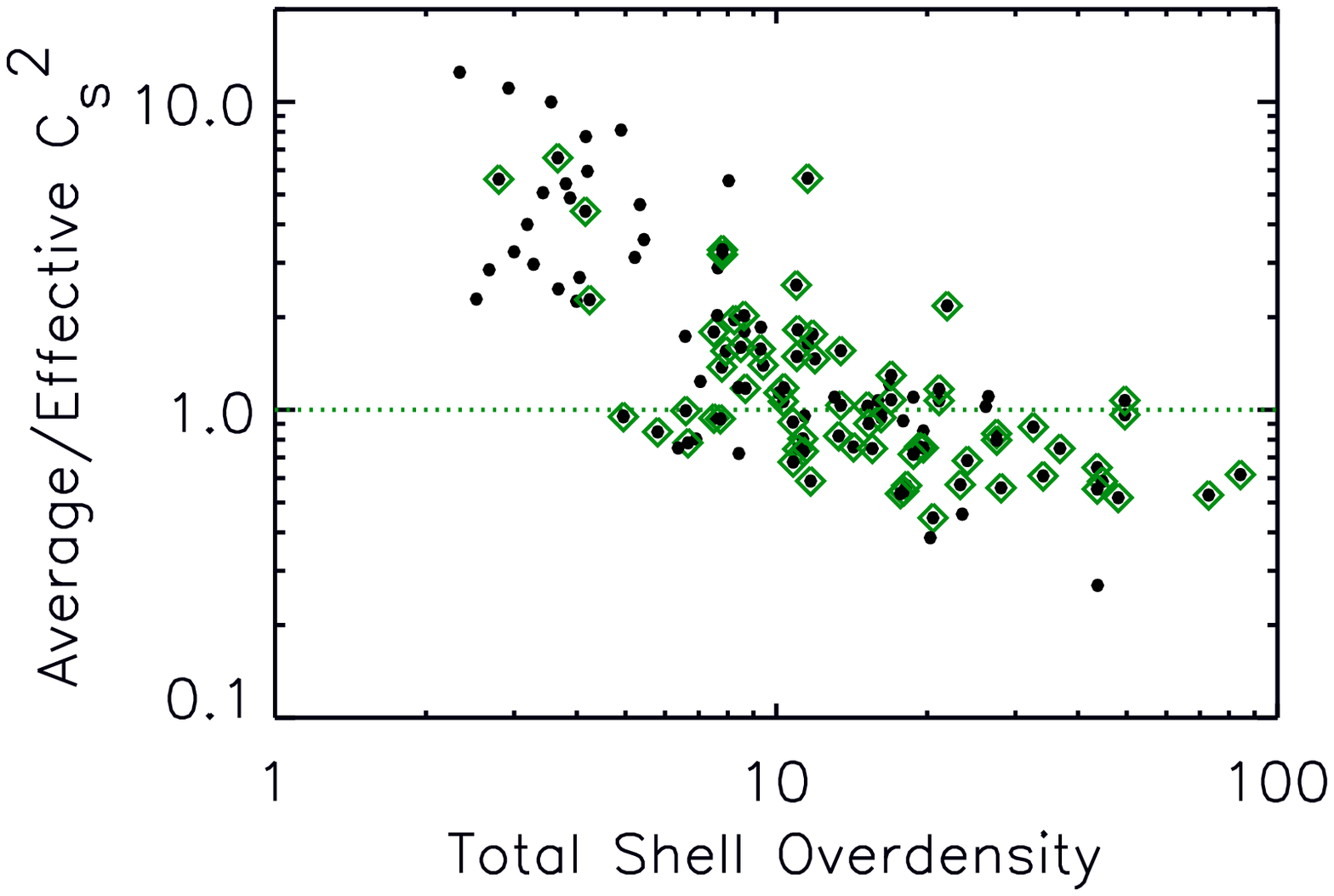}
    \end{center}
    \caption[1]{
    \label{shell_ssq_rat}
    Ratio of the square of the average actual sound speed to the 
    that of the effective one as a function of the
    total overdensity of the parent shell.
    All of the points
    represent isothermal-like segments.  Those segments
    having gas fractions
    greater than one half are enclosed in green diamonds.
    Dotted green line marks a ratio of one.
    }
    \end{figure}
}

\newcommand{\cmdgfscatt}{
    \begin{figure}
    \begin{center}
    \leavevmode
    \includegraphics[bb=50 0 504 360,scale=0.55]{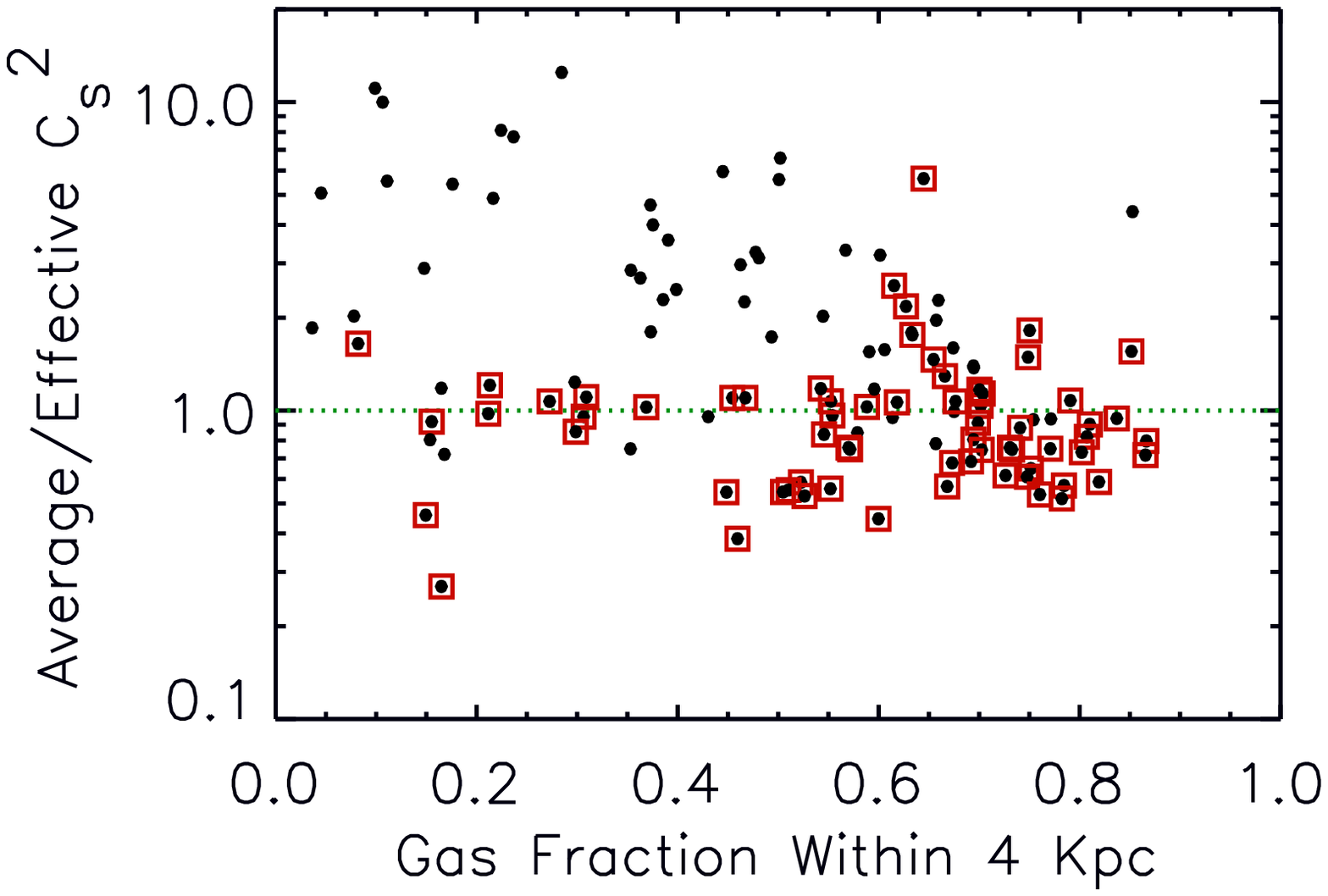}
    \end{center}
    \caption[1]{
    \label{gf_scatt}
    Ratio of the square of the average actual sound speed to the 
    that of the effective one as a function of the gas fraction
    within four kpc of the segment axis.
    All of the points represent isothermal-like segments.  Those that
    have parent shell overdensities greater than ten are enclosed in
    red boxes.
    Dotted, green line marks a ratio of one.
    }
    \end{figure}
}

\newcommand{\cmdrn}{
    \begin{figure}
    \begin{center}
    \leavevmode
    \includegraphics[bb=50 0 504 360,scale=0.5]{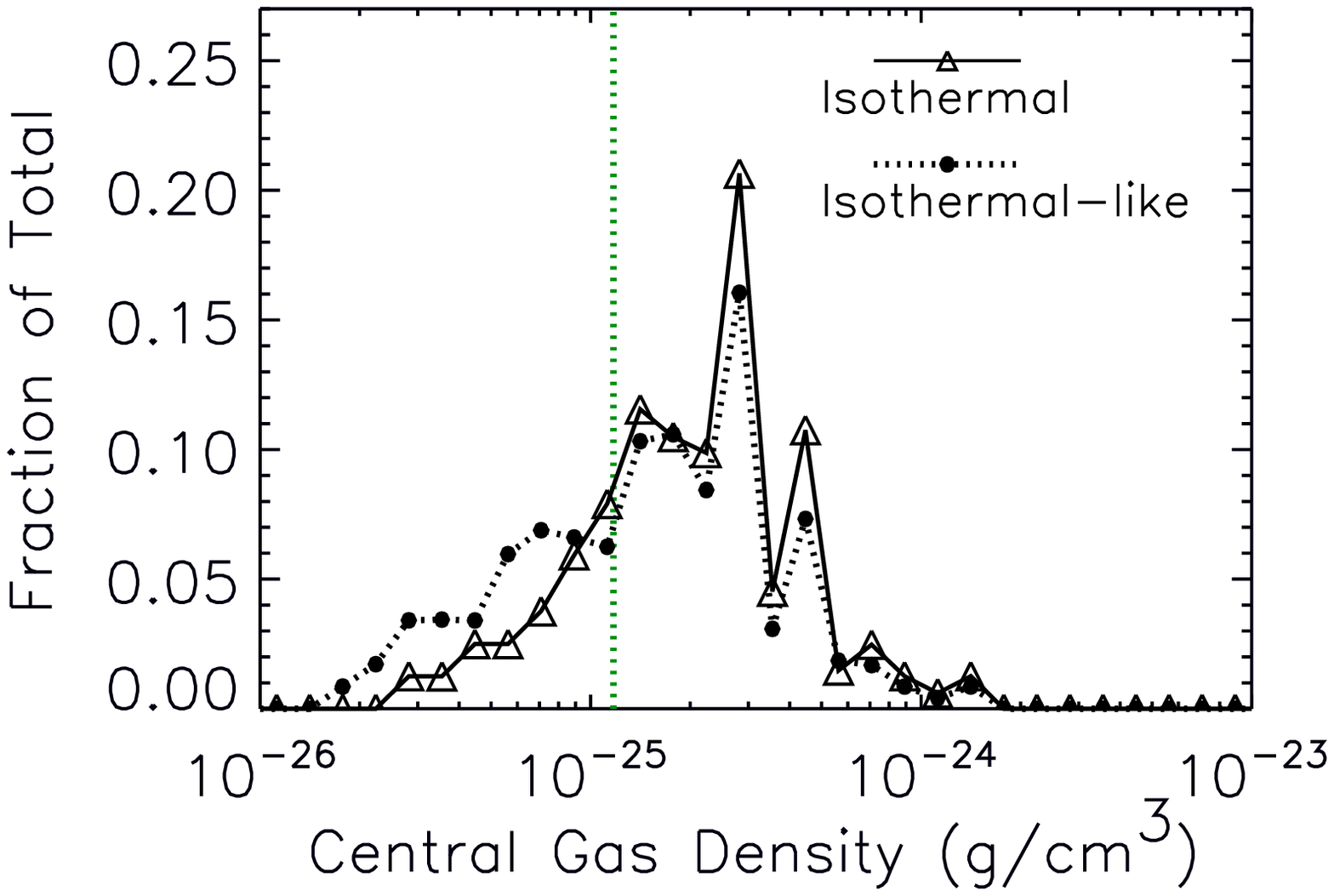}
    \end{center}
    \caption[1]{
    \label{rn}
    \bold{Histogram of computed central gas density 
    of the isothermal and isothermal-like segments.}  
    \bold{The bin size is 0.1 on a logarithmic scale.}
    The abscissa is the gas density in grams per cubic centimeter.
    The dotted, green line shows the gas density that would
    correspond, by itself without dark matter, to a cosmic 
    overdensity of 200.  Isothermal segments are indicated by the
    solid line with triangles.  Isothermal-like segments by 
    the dotted line
    with filled circles.
    }
    \end{figure}
}

\newcommand{\cmdallscatt}{
    \begin{figure}
    \begin{center}
    \leavevmode
    \includegraphics[bb=50 0 504 360,scale=0.55]{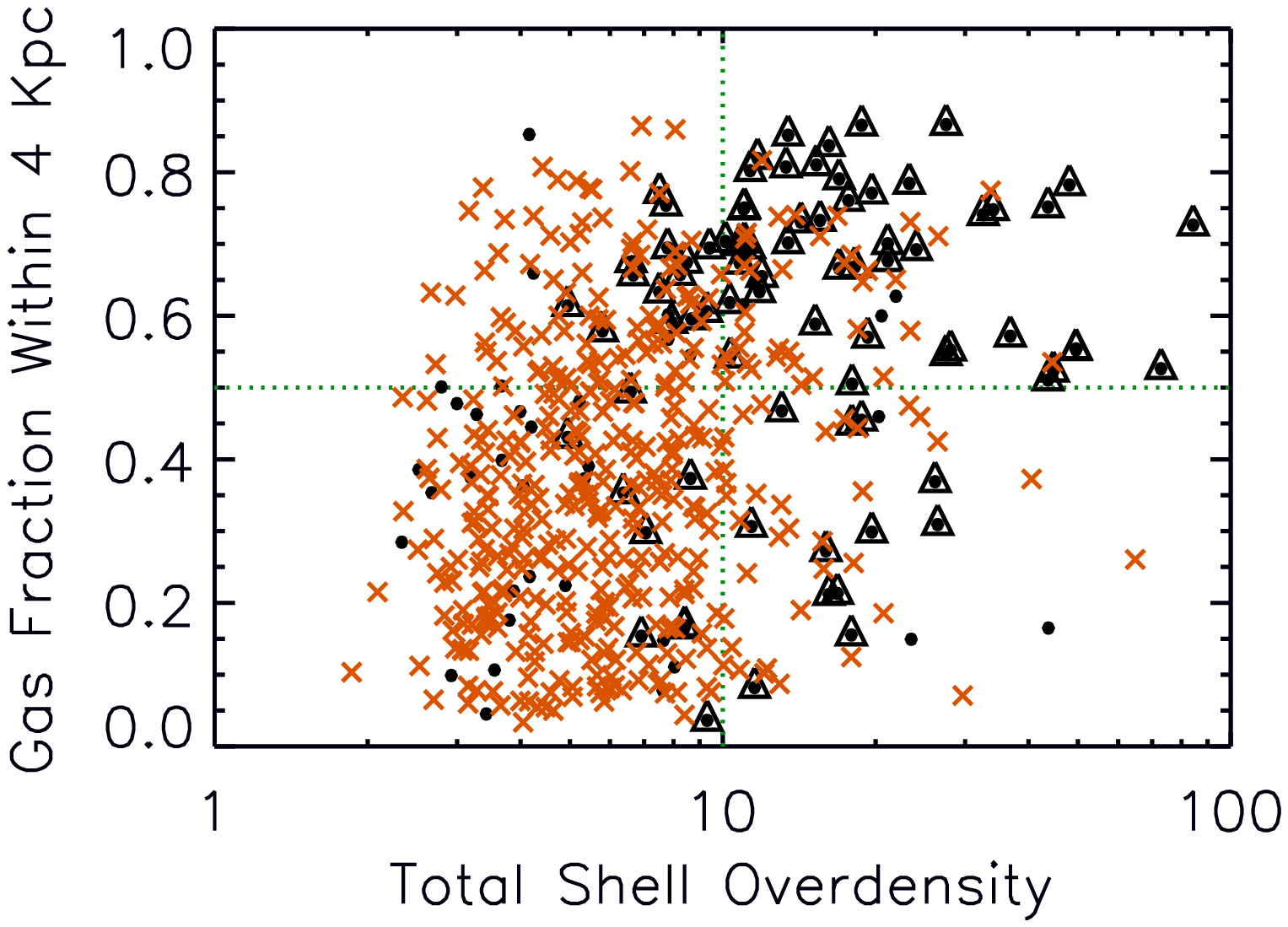}
    \end{center}
    \caption[1]{
    \label{all_scatt}
    Gas fraction within 4 kpc versus total parent
    shell overdensity.  
    Filled black circles are the isothermal-like segments.
    Those of the isothermal-like segments that are
    isothermal are enclosed in black
    triangles.
    Orange X's show segments that are
    not isothermal-like.
    }
    \end{figure}
}

\newcommand{\cmdhone}{
    \begin{figure}
    \begin{center}
    \leavevmode
    \includegraphics[bb=50 0 504 360,scale=0.5]{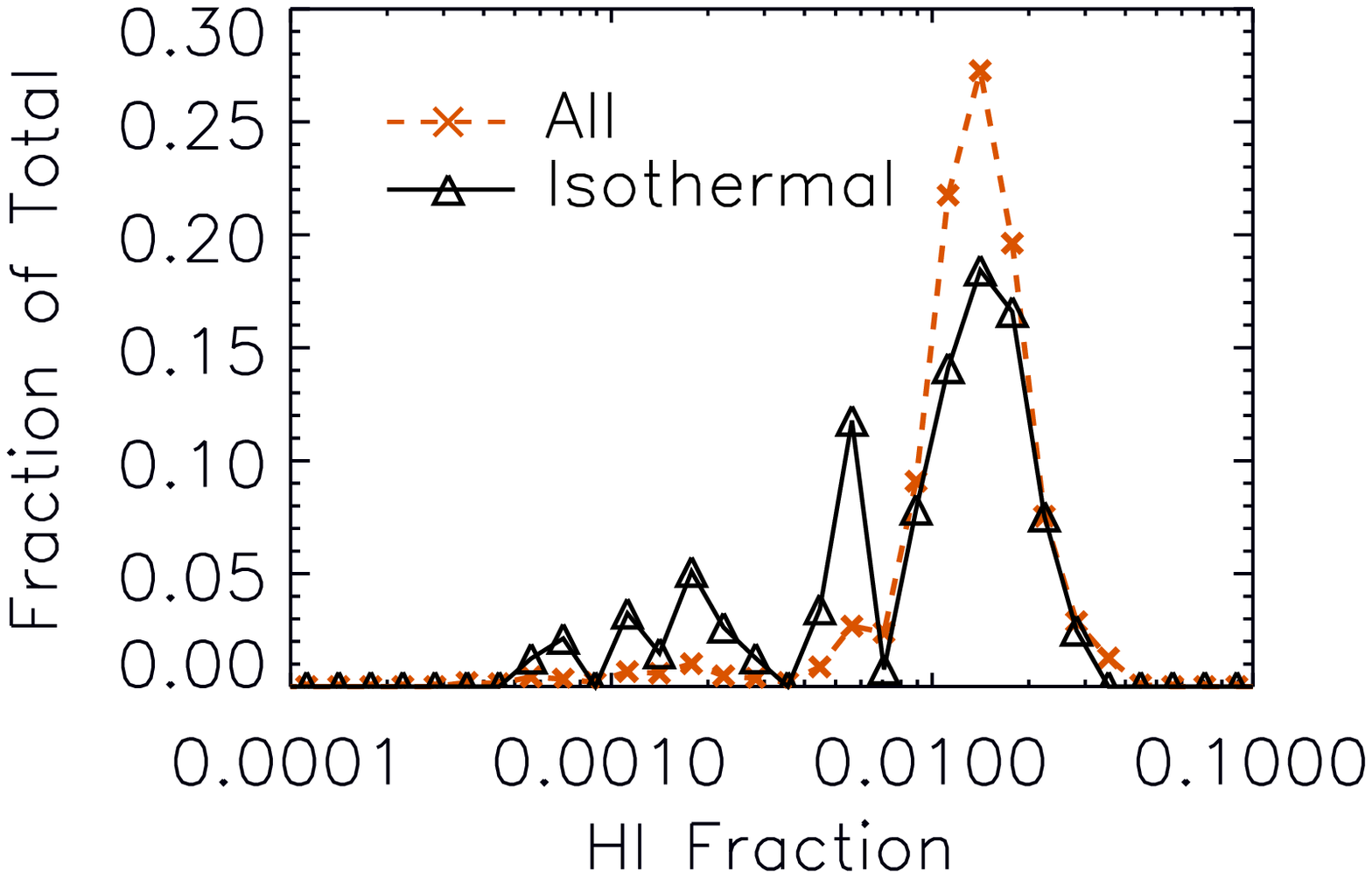}
    \end{center}
    \caption[1]{
    \label{hone}
    \bold{Histogram of HI fractions of hydrogen in segments.}  
    \bold{The bin size is 0.1 on a logarithmic scale.}
    The HI fraction was averaged by mass for the region within \bold{7}
    kpc of the axis of the segment.  Dashed, orange line with X's
    represents all \bold{segments.}  Solid, black line with
    triangles the isothermal ones.
    }
    \end{figure}
}

\newcommand{\cmdssqrat}{
    \begin{figure}
    \begin{center}
    \leavevmode
    \includegraphics[bb=50 0 504 360,scale=0.5]{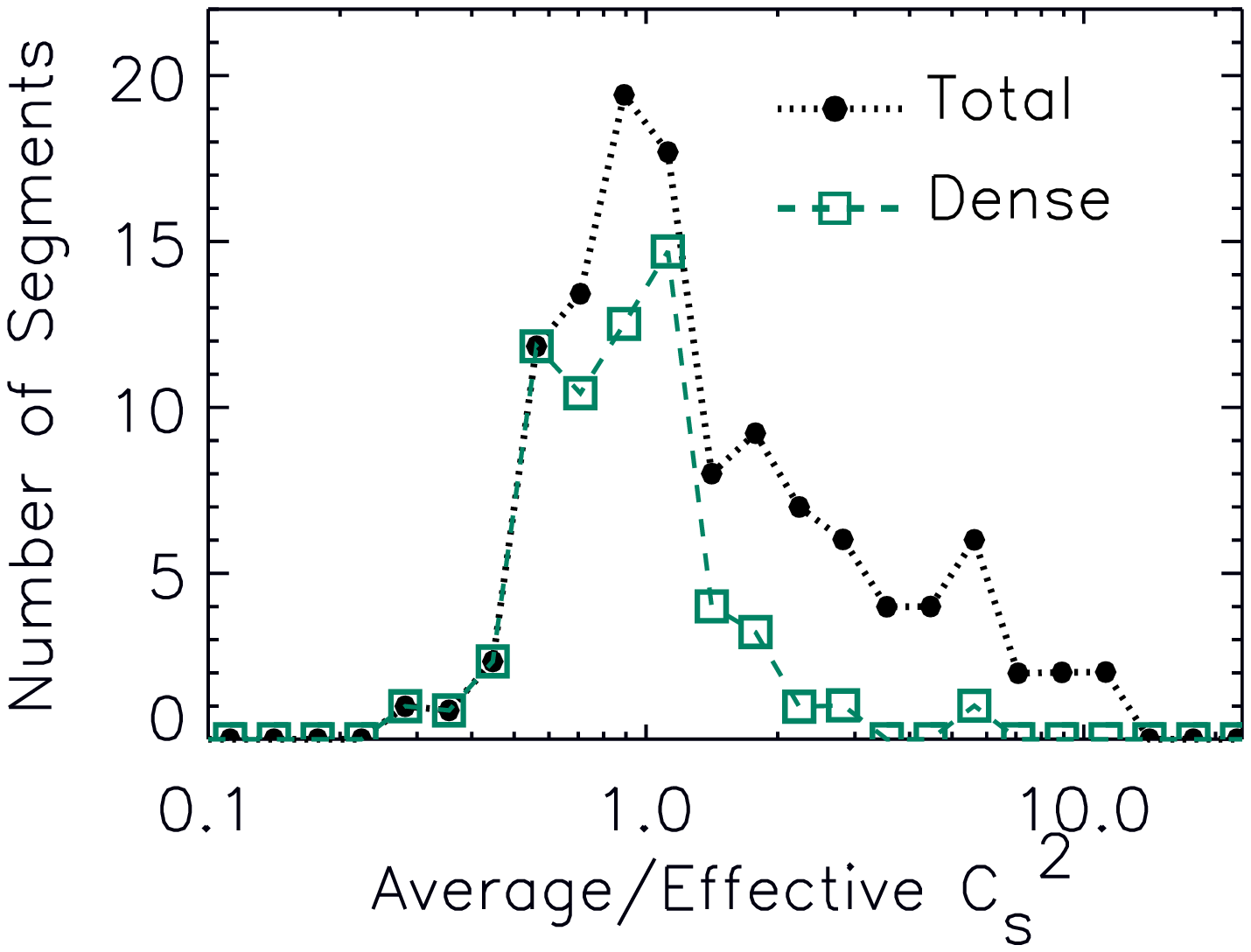}
    \end{center}
    \caption[1]{
    \label{ssq_rat}
    \bold{Histogram of}the ratio of the squares of the actual and
    effective sound speeds
    for the isothermal-like segments 
    (dotted, black line with filled circles)
    and for the subset from parent shells having a total 
    overdensity greater
    than ten (dashed, blue line with squares).  \bold{The bin size is
    0.1 on a logarithmic scale.}
    Overlaps have
    been eliminated for this figure as described in Section~\ref{eval}.
    }
    \end{figure}
}

\newcommand{\cmdshellfinal}{
    \begin{figure}
    \begin{center}
    \leavevmode
    \includegraphics[bb=50 0 504 360,scale=0.45]{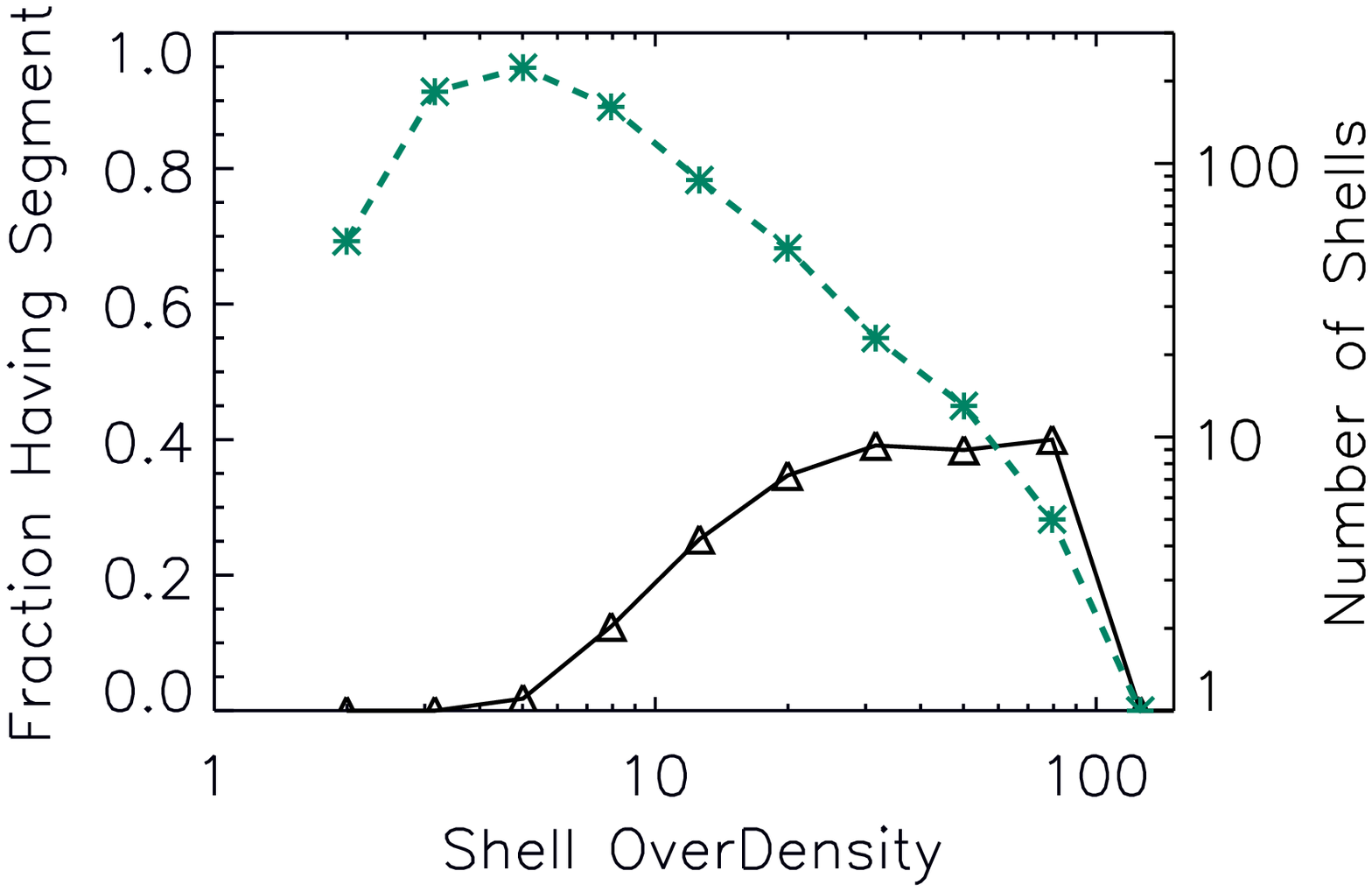}
    \end{center}
    \caption[1]{
    \label{shell_final}
    \bold{Histogram of} occupation of shells by isothermal cylinders.
    \bold{The bin size is 0.2 on a logarithmic scale.}
    The dashed, blue
    line with stars, referring to the ordinate on the right,
    is a histogram of the overdensities of the
    shells in our study.  The solid, black line with triangles,
    referring to the ordinate on the left, shows
    the fraction of shells having at least one isothermal segment.
    }
    \end{figure}
}

\newcommand{\cmdquality}{
    \begin{figure*}
    \begin{minipage}{175mm}
    \includegraphics[scale=0.3]{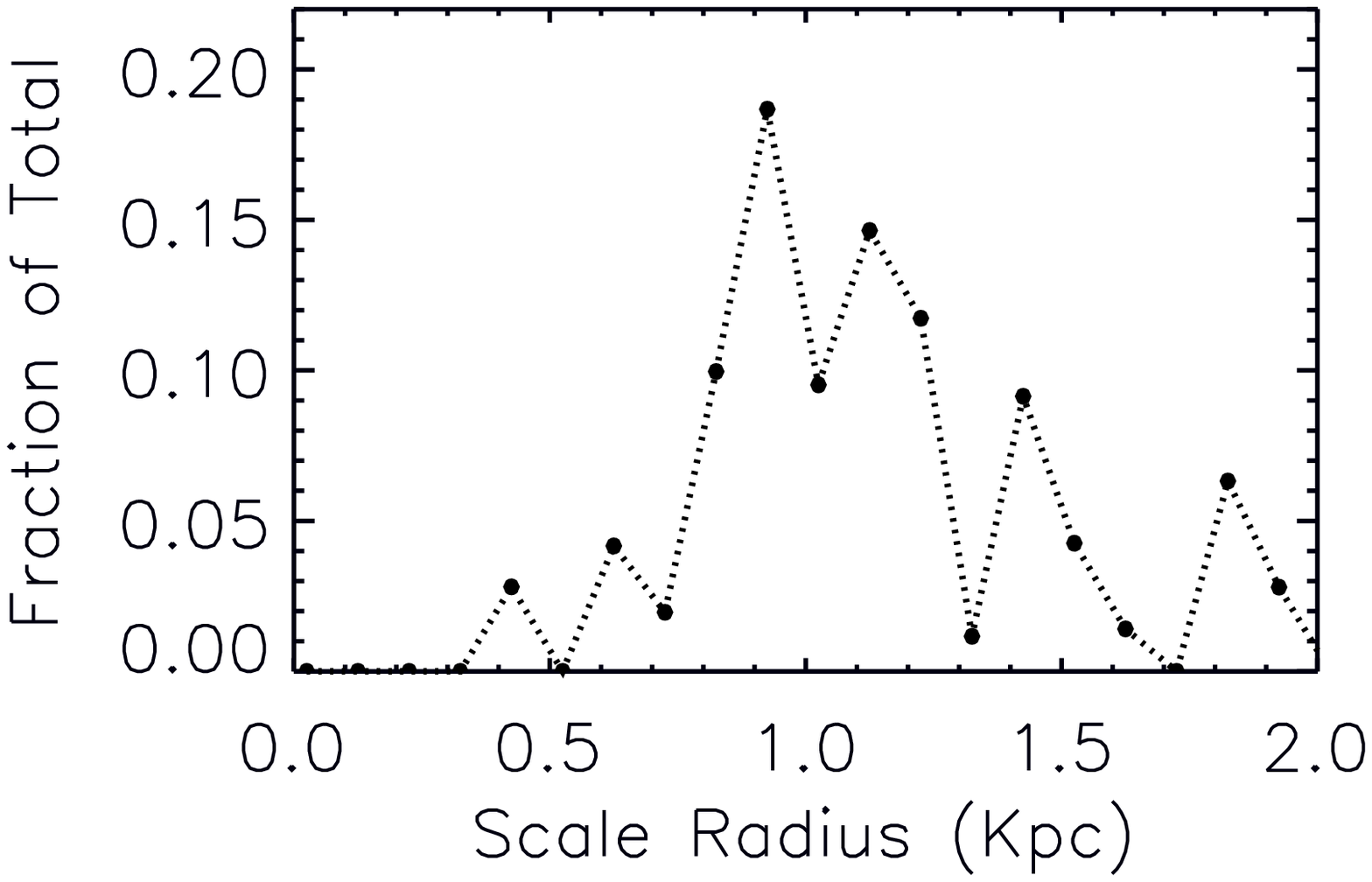}
    \includegraphics[scale=0.3]{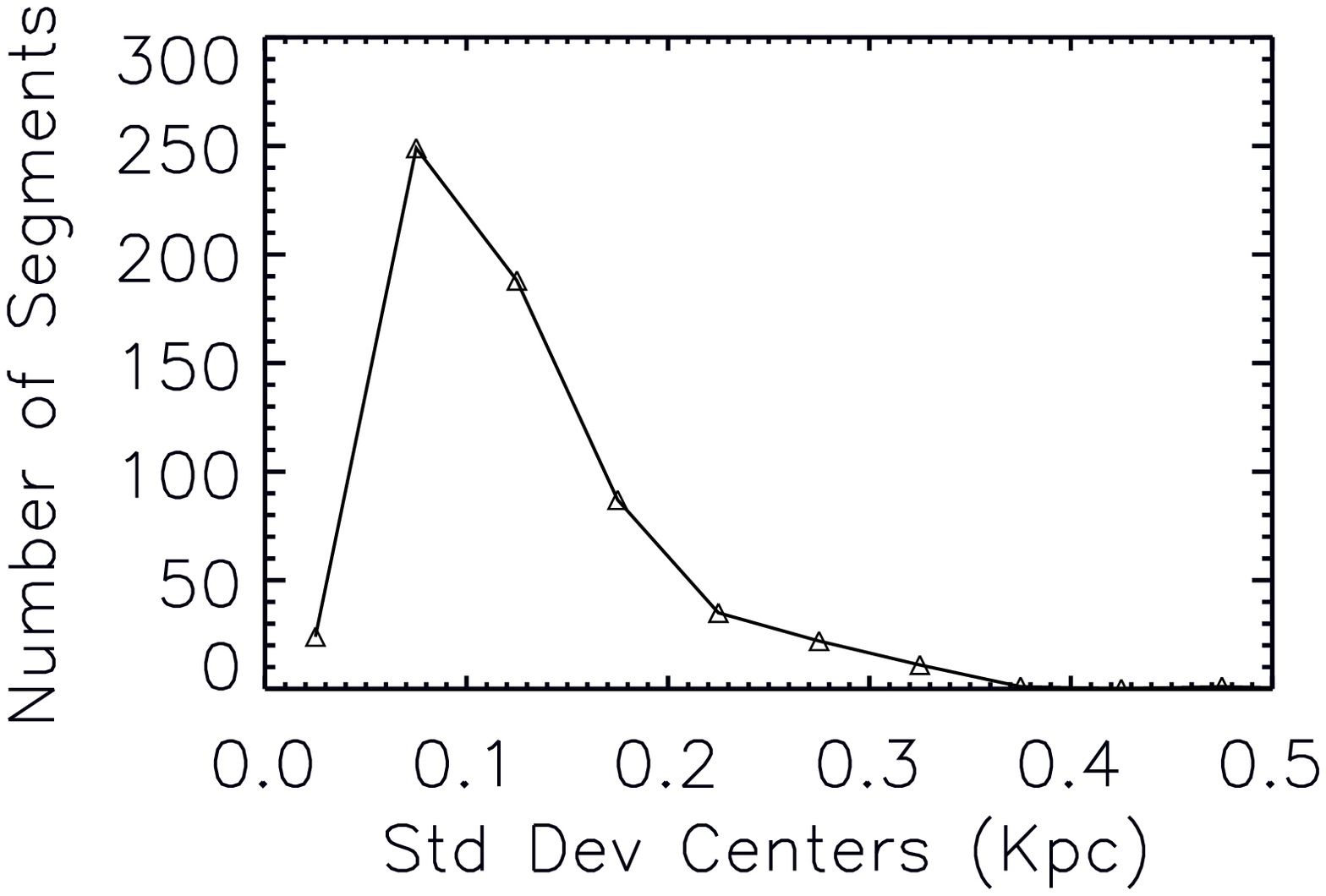}
    \includegraphics[scale=0.3]{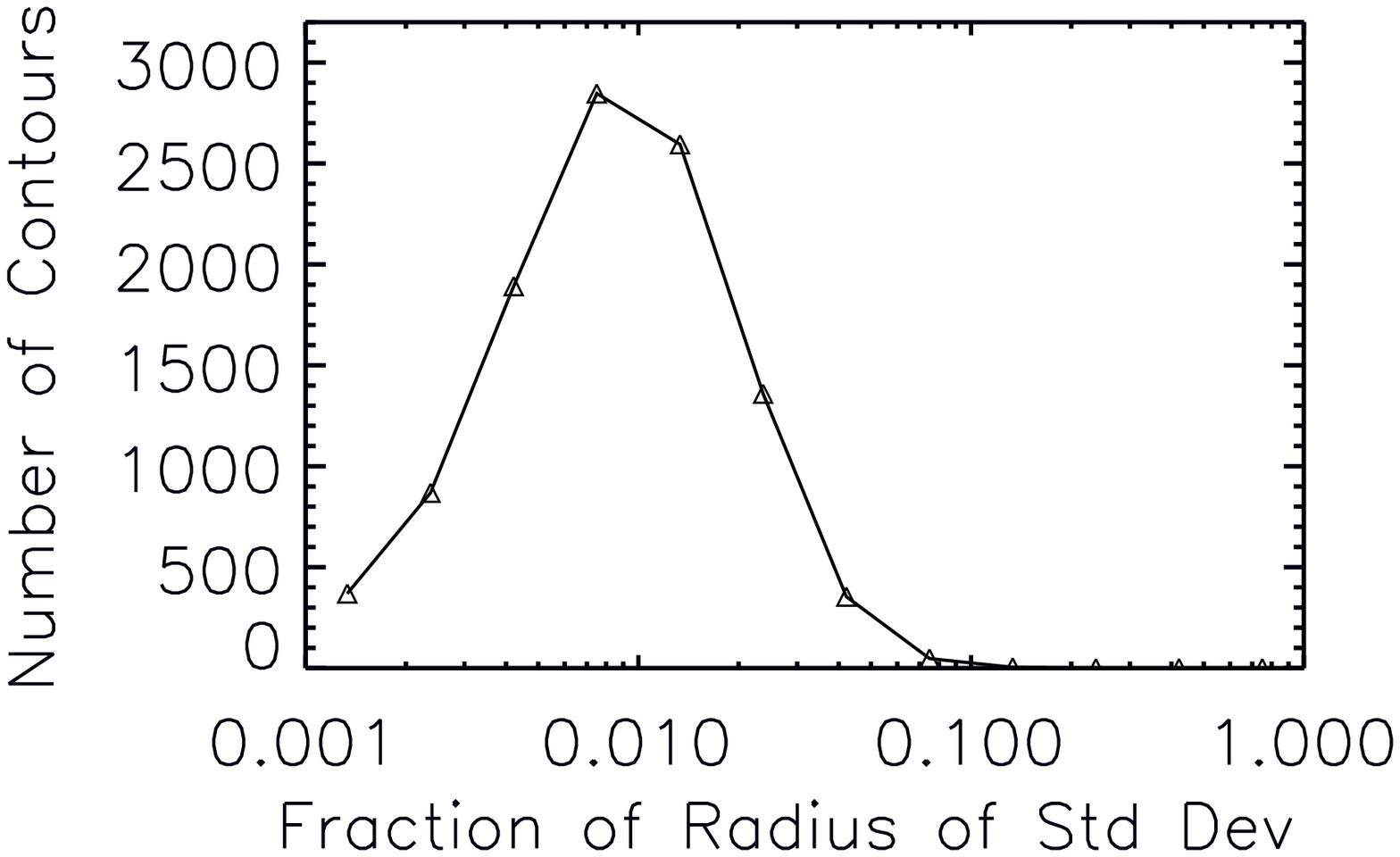}
    \caption[1]{
    \label{quality}
    Left:  Histogram of scale radii for the isothermal-like
    segments, that is, those segments having an isothermal profile for
    some effective temperature. \bold{The bin size is 0.1kpc.}

    Center:  Test of whether contour centers coincide.  
    Shown is a histogram of the
    standard deviation of contour centers from mean of the contour
    centers for each of the segments. \bold{The bin size
    is 0.05 kpc.}
    Included in the figure are
    all contours from the fitted region corresponding to the best fit.
    Segments are 7.24 kpc in length
    and are analyzed individually.

    Right:  Test of the straightness and cylindrical symmetry of the
    individual segments.  Shown is a histogram of the
    standard deviation of the mean of the radius of the
    individual points that make up each contour.  \bold{The bin size
    is 0.25 on a logarithmic scale.}
    The standard deviation is expressed as a fraction of the 
    mean radius of the contour.   Included are all segment
    contours from the fitted region corresponding to the best fit.
    }
    \end{minipage}
    \end{figure*}
}

\newcommand{\cmdshock}{
    \begin{figure}
    \begin{center}
    \leavevmode
    \includegraphics[bb=50 0 504 360,scale=0.55]{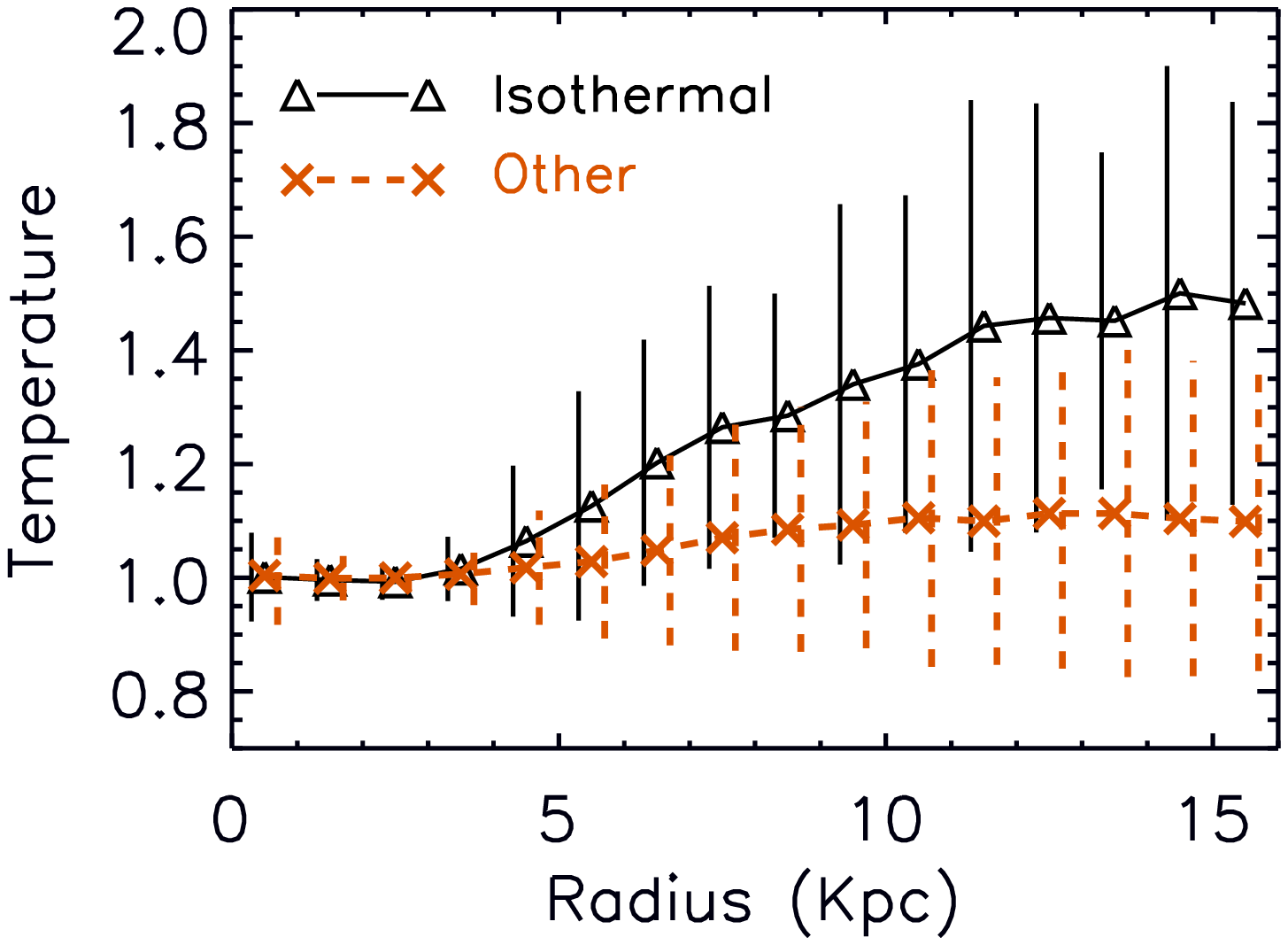}
    \end{center}
    \caption[1]{
    \label{shock}
    \bold{Cylindrical temperature profiles of filament segments.
    Shown is the average temperature as a function of
    distance from the segment axis.  Each segment temperature 
    in the average is normalized to the average temperature
    of that segment enclosed within a radius of 4 kpc.
    The vertical lines are standard deviations of the individual
    normalized temperatures (not the standard deviation of the mean).
    Black line with triangles represents the isothermal segments.
    Orange line with X's the remaining segments, that do not
    meet the requirements to be isothermal.}
    }
    \end{figure}
}

\newcommand{\cmdten}{
    \begin{figure}
    \begin{center}
    \leavevmode
    \includegraphics[bb=50 0 504 360,scale=0.45]{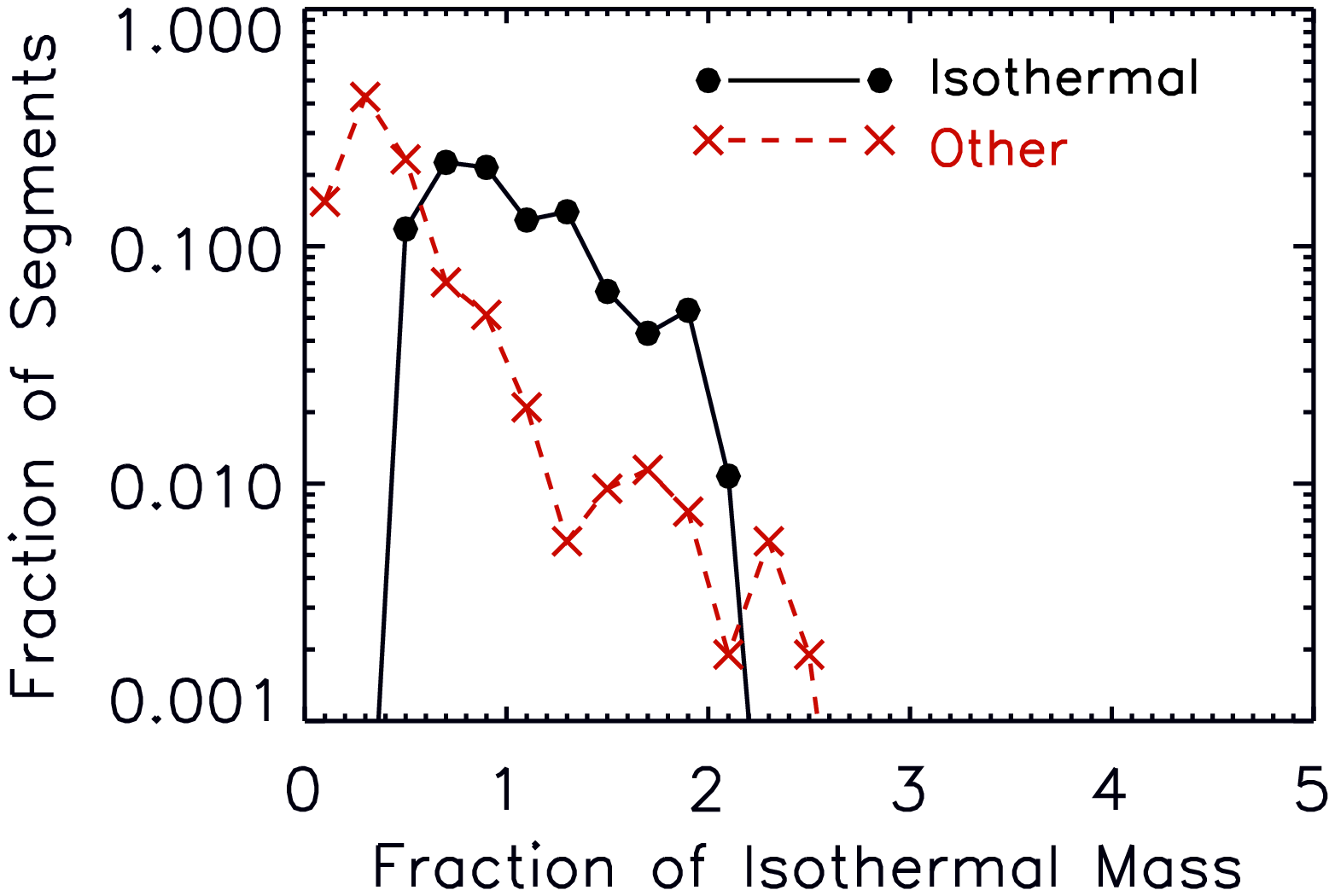}
    \end{center}
    \caption[1]{
    \label{ten}
    \bold{Gas associated with segments.  Histogram showing 
    the ratio of the mass of gas
    within 10 kpc of the segment axis to the total gas mass that an
    isothermal cylinder of the corresponding temperature and
    ionization state would have.  Bin size is 0.2.
    The black line with filled circles represents the isothermal
    segments and the red, dashed line with X's the remaining
    segments.}
    }
    \end{figure}
}

\newcommand{\cmdvelz}{
    \begin{figure}
    \begin{center}
    \leavevmode
    \includegraphics[bb=50 0 504 360,scale=0.55]{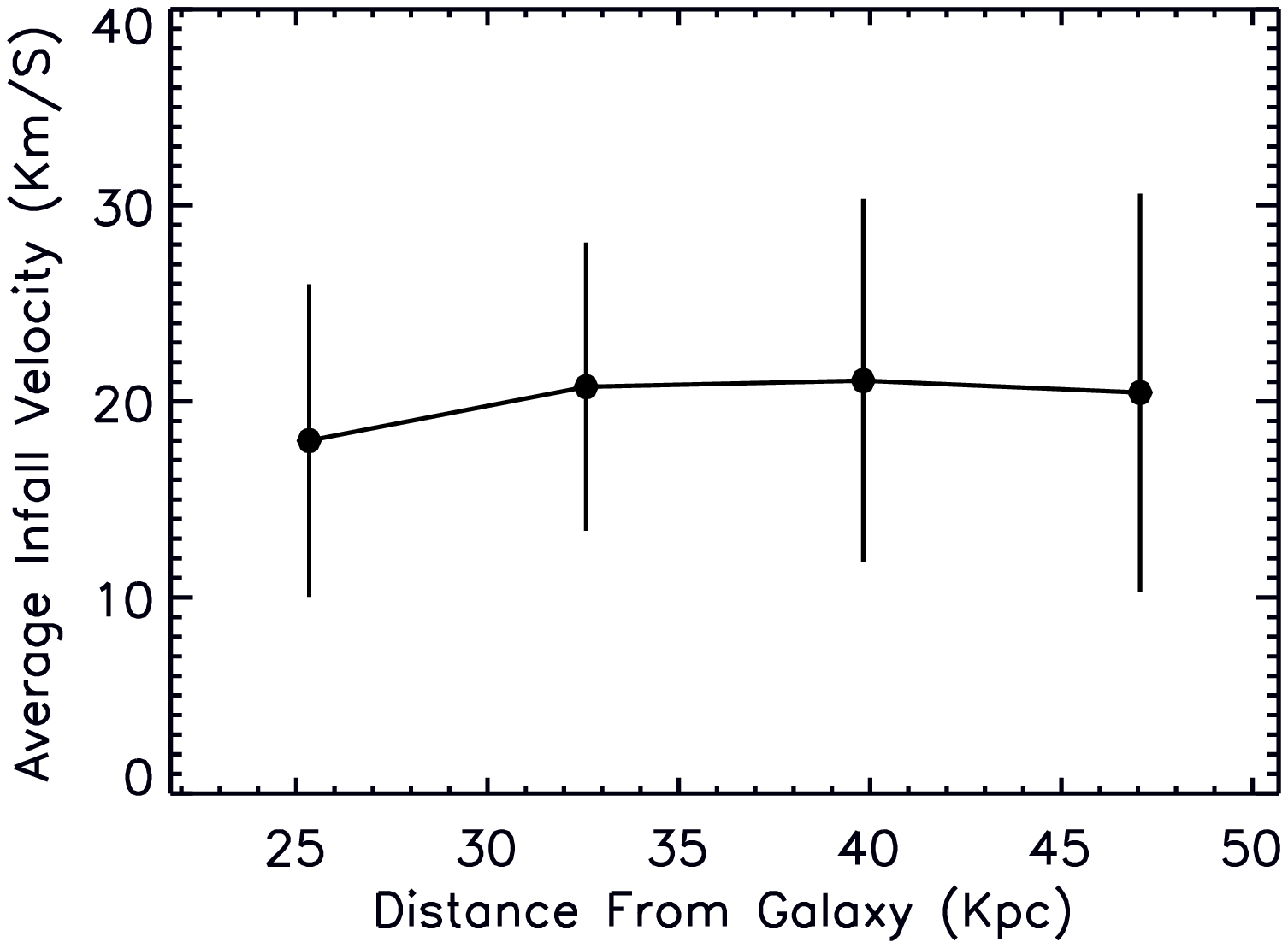}
    \end{center}
    \caption[1]{
    \label{velz}
    \bold{Average velocity of gas infall toward the 
    galaxy as a function
    of distance of segment from galaxy.  Shown are average 
    peculiar velocities
    of gas particles within 4 kpc of the cylinder axis after
    subtracting out the average peculiar velocity of the 
    stars at the
    center of the galaxy.  Vertical lines show twice the standard
    deviation of the mean.}
    }
    \end{figure}
}

\newcommand{\cmdtable}{
    \begin{table*}
    \begin{minipage}{80mm}
    \caption{Numbers of Segments.  Fractions are by numbers
    of segments or by gas mass (in parentheses).  Overlaps
    have been eliminated as described in section~\ref{eval}}.
    \label{the_table}
    \begin{tabular}{lcccc}
Criterion & Total & Isothermal & Isothermal/Total \\
None & 540 & 80 & 0.15 (0.32) \\
Density & 136 & 57 & 0.42 (0.52) \\
Density and Gas Fraction & 86 & 44 & 0.52 (0.63) \\ 
    \end{tabular}
    \end{minipage}
    \end{table*}
}

\begin{document}

\maketitle

\begin{abstract}

Using a cosmological simulation at redshift 5, we find that the 
baryon-rich cores of intergalactic filaments \bold{extending outward}
from galaxies commonly
form isothermal gas cylinders \bold{in regions favorable to their
formation.} 
The central gas density is typically about 500
times the cosmic mean total density, and the 
temperature is
typically 1-2 times $10^{4}\dim{K}$, just above the
Lyman alpha cooling floor.  These findings argue that
the hydrodynamic properties of 
the gas are more important than the dark matter in
determining this structure.  
\bold{It is noteworthy that the temperature and ionization 
state of the gas
completely determine a finite total mass per unit length of an
isothermal cylinder.  Our findings may have implications for 
understanding the ``cold mode'' mechanism of gas transport into
galaxies.}
\end{abstract}


\section{Introduction}
\label{intro}

The standard model of cosmology emphasizes the role of the
gravitational field of the dark matter in structure
formation.  In this picture, the dark
matter determines the overall geometry and rate
of growth of structure.

While analyzing a cosmological simulation at redshift 5, we found that
on scales of a few proper kiloparsecs intergalactic dark matter 
and baryons form qualitatively different structures \citep{harford08}.  
In contrast to the 
dark matter, which tends to occur in many small, quasi-spherical
clumps, the baryons occur in a framework of thin, smooth rods 
that form backbones connecting the large, dark-matter dominated
galaxies.  

These gaseous filaments likely correspond to the filaments
described by others that
mediate the ``cold mode'' mechanism of gas transport into
galaxies \citep{birnboim03,binney04,katz03,keres05,dekel06,ocvirk08,
keres09b,dekel09d,brooks09}. 
\bold{In contrast to the classical picture in which incoming gas 
is shock-heated
to the virial temperature of the dark matter halo, gas accreted by this
mechanism is never shock-heated, but rather moves directly to the
center of the galaxy along relatively cool filaments.  
This mechanism predominates for \bolder{haloes} less massive than 
about  $10^{11.5}\dim{M}_\odot$
(dark matter, gas, and stars combined).  These \bolder{haloes}
appear to be too small to sustain a shock \citep{birnboim03}.  
If the filaments we have studied
are indeed conduits for such accretion, then their 
structure might have
ramifications for understanding how galaxies gain the gas necessary
for the varied star formation histories that are observed.}

We reasoned that these contrasting distributions of gas and 
dark matter
result from hydrodynamic effects:  pressure forces
retard the gas as it moves toward the axis of a filament, 
in contrast to the dark matter which
can pass freely through subject only to gravitational forces.  
The gas would be expected to accumulate along the axis of the filament
until the pressure forces are sufficient to counteract
the gravitational field.
In this paper we explore the hypothesis that the gas in the 
baryonic core forms a self-gravitating, isothermal cylinder
in hydrostatic equilibrium,
whose structure is determined  primarily 
by the gravitational and hydrodynamic properties of the gas.  

Many different versions of an isothermal cylinder are 
possible with different
degrees of concentration of the gas, but they all share the
interesting property that the mass per unit length of the 
gas \bold{is finite and}
depends only on the temperature and ionization state
\citep{stod63,ostriker64}.  We suggest that
our findings may place constraints upon models for
the movement of gas along filaments.

The importance of hydrodynamic forces for the filament structure 
supports a hydrodynamic mechanism of origin from isothermal sheets 
as first proposed 
by \citet{schmidburgk67}.

\bold{The plan of the paper is as follows:  Section~\ref{ration}
summarizes our previous work on intergalactic filaments, which
provides the rationale for the present study.  The important
properties of the infinite, self-gravitating, isothermal cylinder
are then presented, followed by the important features of our
cosmological simulation.
The results of our study are presented in
Section~\ref{results}.  In Section~\ref{methods} we first
review the details of the simulation. Then the  details of
the procedures for selecting filaments, curve fitting, and  
the elimination of overlaps are presented.
Finally, Section~\ref{discussion} 
summarizes our conclusions and discusses some broader
implications.}

\cmdhighway

\section{Background and Rationale}\label{ration}
In Paper I \citep{harford08}
we studied the distributions
of dark matter and baryons in a detailed cosmological
simulation run to a redshift of 5.1.  At this
redshift
we found that the distributions of dark matter
and baryons begin to diverge in an interesting way
at a scale less than
40 kpc comoving.  Dividing the simulation box
into cubical grid cells of this dimension, we found
that, among cells with a total overdensity of $10^{1.75}$
or greater, about 5\% have a baryonic fraction 
exceeding twice the cosmic mean.  These enriched cells,
in contrast to their unenriched peers, form a network
of long, thin filaments that connect the largest
galaxies.  We studied the filaments \bold{that issue} from the
ten largest galaxies
in detail and found that, in addition to being baryon-rich,
they \bold{extend outward} in approximately
straight lines from the centers of the galaxies.
\bold{We decided to study the structure of these filaments
in more detail.}

\bold{Our working hypothesis is that the filaments form
self-gravitating, isothermal cylinders in hydrostatic
equilibrium.  This hypothesis
is motivated by our finding that, for many of
the filaments, the inner core, within four proper kpc
of the filament axis, is predominately gas.  
\bold{Figure~\ref{highway}
shows the typical distributions of gas and dark matter about
a filament. Each image shows actual simulation
particles in a sub-box of the simulation.  The top
image shows just the gas particles, while the bottom
image shows just the dark matter particles.  Clearly
the gas and dark matter have very different
structures.}

Since our maximum \bolder{halo}
mass (dark matter, gas, and stars combined) is 
$10^{11.07}\dim{M}_\odot$, shocks are generally 
not expected to form.  Consistent with this expectation,
the gas bounding the filaments has a temperature of only
a few times $10^{4}\dim{K}$.}

The infinite, self-gravitating, isothermal gas cylinder in hydrodynamical
equilibrium has been described mathematically
by \citet{stod63} and \citet{ostriker64}.  The density profile is given by

\begin{equation}
\label{rho_eq}
\rho = \rho_0 \frac{1}{(1+\tfrac{1}{8}\xi^2)^2}
\end{equation}
where $\rho_0$ is the central density along 
the axis of the cylinder and
\begin{equation}
\label{xi_eq}
\xi = \frac{r}{r_s}
\end{equation}
where $r$ is the radial distance and $r_s$ the scale
radius
\begin{equation}
r_s = \frac{c_s}{\sqrtsign{4\pi G\rho_0}}.
\end{equation}
In this expression $c_s$ is the isothermal speed of sound
and $G$ is the gravitational constant.
The speed of sound depends only upon the temperature
$T$ and the mean molecular weight $\mu$
\begin{equation}
c_s = \sqrtsign{\frac{kT}{\mu m_H}}
\end{equation}
where $k$ is the Boltzmann constant and $m_H$ the mass
of the hydrogen atom.

The total mass per unit length of an isothermal cylinder,
\bold{obtained by integrating equation \eqref{rho_eq},
is finite and given by}
\begin{equation}
\frac{2 c_s^{2}}{G}
\end{equation}
where $c_s$ is the \bold{isothermal} sound speed and $G$ is the gravitational
constant.  \bold{This result agrees with the virial theorem
for an infinite cylinder derived by \citet{chandra53}.}

Because of the limited number of simulation particles, it
is convenient to work with the gravitational potential produced
by the distribution of the gas rather than with the density profile
itself.   Dividing the potential by $c_s^2$ gives a dimensionless
variable $\psi$ which takes the simple form
\begin{equation}
\label{psi_eq}
\psi(\xi) = 2\ln(1+\tfrac{1}{8}\xi^2)
\end{equation}
for the isothermal cylinder.

\bold{In this paper we will use the term isothermal-like to refer to
a distribution of gas whose potential satisfies equation \eqref{psi_eq}
for some sound speed.
If the sound speed is equal to the
actual sound speed of the gas (within a factor of two), 
the distribution will also be called isothermal.
A gas distribution inconsistent with \eqref{psi_eq}
will be termed non-isothermal.}

Filaments in a cosmological simulation
are of course messier than this idealization.
They may be bent, truncated, or disturbed
in various ways by galaxies, small dark matter haloes,
or other filaments.  Thus at best it can be shown only that
relatively undisturbed filaments emanating from galaxies are
commonly reasonably well approximated as isothermal cylinders.
Despite this limitation, such a finding would be interesting
because it would suggest that the dynamics of filaments at scales
of a few kpc at redshift 5 is dominated by gaseous processes
rather than by the gravity of dark matter. 

In anticipation of considerable variability,
we expanded our study to include the filaments around the 
200 largest \bolder{haloes} \bold{in our $8h^{-1} \unit{Mpc}$ comoving
simulation box.}   These \bolder{haloes}
range in total mass \bold{(gas, dark matter, and stars)}
from
$10^{9.70}\dim{M}_\odot$ to $10^{11.07}\dim{M}_\odot$, 
\bold{in the mass range for shockless cold mode
accretion.}  \bolder{The mass of a halo is defined as the mass 
gravitationally bound to it as opposed to the mass within the virial
radius.  The differences are not large.}

We used the same simulation at the same redshift as in Paper I.
A feature that distinguishes this simulation from many others in the
literature is that the ionizing radiation is generated
\bold{directly} from the simulated star formation rather
than added as a uniform background.  Detailed chemistry and
radiative transfer computations are included.  Resultant luminosity
functions are reasonable \citep{harford03,harford07}.  Although the
history of reionization remains controversial, the process within the
simulation is at least consistent 
with the spectra of high-redshift quasars 
\citep{gnedin_fan06}.  The importance of the latter is underscored by
the finding that the Jeans length changes by more than an
order of magnitude over the course of reionization 
(see \citealt{gnedin_filt03} and references therein).

\cmdscheme

In what follows, all distances are proper unless specified
otherwise.  All histograms are shown with the plotted point at the
center of the bin.

\section{Results}\label{results}

This section presents the basic results of our study.
Details of filament selection and methodology are deferred
to Section~\ref{methods}.

To study a large number of galaxies,
we devised an efficient, objective way to identify filaments.
In Paper I we found that the
baryon-rich filaments extend from the
centers of the galaxies in a predominantly radial direction,
consistent with their being conduits for gas accretion onto
galaxies.  From this perspective, we take as our defining feature
of a filament a long, straight
rod of gas protruding from the center of a galaxy.
To implement this definition
we used the HealPix \citep{gorski05} partition of the volume 
of a sphere centered on each galaxy.
This method divides the region completely into volume
elements of equal solid angle issuing from the center. We
reasoned that the volume elements
containing the most gas probably represent directions of
filaments.

\cmdgforig
\cmdtmptr
\cmdhone
\cmdshock

To allow for the possibility that the filament
structure might vary along its length, we divided filaments into
segments \bold{7.24 kpc long for individual analysis.}
Each angular volume element is divided into four radial
parts corresponding to four spherical shells bracketed by 
five spheres at \bold{radii of} 3, 4, 5, 6, and
7 radial units (7.24 kpc) from the center of the galaxy.
\bold{The spherical shell 
in which a segment resides will be termed its parent shell,
and its total overdensity will largely
determine its ability to host an isothermal structure.}

The minimum distance from the galaxy center is thus 22 kpc,
outside the virial radius of all but a few of the larger
galaxies.  The virial radii
\footnote{The virial radius was taken
to be the radius enclosing an average total overdensity of 200 times
the cosmic mean \bold{and was computed directly from the simulation.  
The total overdensity includes gas, dark matter,
and stars.}} range from 28 kpc for
the largest galaxy down to 6 kpc. Details of the
selection and analysis of filament segments are given
in Section~\ref{find}.

\bold{An illustrative example is shown in Figure~\ref{scheme}.
Each of the four images represents the same spherical volume 
of radius 51 kpc centered on one galaxy.
The top left image shows only the gas particles in  this region,
while the top right shows only the dark matter particles.  
The two gas filaments issuing from the center of the galaxy
are well delineated.
This example exemplifies our working hypothesis that the gas has a
distinctive structure unlike that of the dark matter.  
The bottom left image shows gas particles in the 
individual filament segments
that we analyzed, colored in order moving outward black, green,
red, and blue.  The remaining gas particles are shown in yellow to
provide contrast.  The bottom right image shows in black
the segments extending upward, which were deemed to be 
isothermal, while the
red ones extending downward were not.}

The segments belong to 
a small number of major strings of galaxies, 
resembling the
filaments formed by the baryon-rich grid cells in Paper I.

Figure~\ref{gf_orig} shows that 
the cores of the majority of the segments
are enriched in gas above the cosmic 
mean, consistent with the findings of Paper I.
The subset of segments that we find to be isothermal
is even more gas-rich.
In the present paper we use the gas fraction
rather than the baryonic fraction \bold{(gas and stars)} used in Paper I.
This convention was chosen
because the stellar particles in the simulation,
like the dark matter ones, are treated
as collisionless, in constrast to the
hydrodynamic treatment of the gas.  The contribution of stars 
to the baryonic fraction is generally small.

The segments
have average temperatures
of about 12000 to 14000 K, just above the Lyman alpha cooling
floor (Figure~\ref{tmptr})
and neutral hydrogen
fractions of about 0.01 to 0.02 (Figure~\ref{hone}).

\bold{The extended temperature profiles shown in Figure~\ref{shock}
show that the segments are bounded mostly by low temperature regions.  
For the isothermal segments the temperature increase in the outer
regions is small, not suggestive of strong shocks or high
pressure confinement of these filaments by high temperature gas.  
The remaining segments
that are not isothermal show essentially no change in temperature.
Thus it seems unlikely that they fail to be isothermal because of
shocks or ram pressure from the surrounding gas.}

\cmdssqrat

The structure of each segment was determined from the cylindrical,
radial profile of the gravitational potential produced
by the distribution of its gas.  Section~\ref{ration} 
described the basic properties of a self-gravitating, isothermal 
cylinder.  Details of the computation and
the fitting of the potential profiles to an isothermal one
are given in
Sections ~\ref{potential} and ~\ref{fitting}.

\cmdshellssqrat
\cmdgfscatt

\bold{From an isothermal potential profile,} an effective sound speed 
can be computed.  This is the sound speed
that the gas would \bold{have if its pressure balanced} its
gravitational field.  The effective sound speed
is then compared to the actual sound 
speed \bold{averaged within 4 kpc of the cylinder axis.}
\bold{Figure~\ref{ssq_rat} shows a histogram of the 
square of the ratio
of the actual sound speed to the effective one for the isothermal-like
segments, that is, those that satisfy equation \eqref{psi_eq}
for some effective sound speed.
There is a pronounced peak at a ratio of one.  The segments whose ratio is
within a factor of two of one will be considered isothermal.}

\cmdtable

\bold{The mass per unit length of an isothermal cylinder depends only
upon the temperature and average molecular weight of the gas.}
For a typical segment with a temperature of
1-2 times $10^{4}\dim{K}$,
this mass is about $4\times10^{8}\dim{M}_\odot$ of gas \bolder{per shell.}
Typically a large galaxy is surrounded by two to three filaments.
The parent shell of a segment would then need a minimum of 
about $10^{9}\dim{M}_\odot$
of total gas to accomodate 2.5 isothermal cylinders.  
Not all of the parent shells even have this much gas.

The scatter plot in Figure~\ref{shell_ssq_rat} compares 
the sound speed ratio to the
total cosmic overdensity of the parent shell.  All of the points
represent isothermal-like segments.
As the shell overdensity falls
below ten the sound speed ratio rises indicating that there is
too little gas to balance the ambient \bold{pressure.}


Figure~\ref{ssq_rat} shows that this density effect accounts for 
most of the the failure of isothermal-like segments to be isothermal. 
The dashed, blue line with squares shows the sound speed
ratio distribution
for segments whose parent shells have a total cosmic overdensity of
at least ten.

Another factor that might be important is the gas fraction of the
segment.  Dark matter might increase the gravitational field enough
to accomodate a higher temperature,
or it might disrupt the isothermal structure altogether.
Figure~\ref{gf_scatt} shows the ratio of the square of the sound
speeds of the isothermal-like segments as a function of gas fraction.
This figure shows that higher gas fractions are more
conducive to isothermal cylinders.

\cmdallscatt

Figure~\ref{all_scatt} combines the two criteria of overdensity
and gas fraction.  The upper right quadrant
of the figure contains those segments
that have both a parent shell of overdensity greater than ten
and a segment gas fraction greater than one half.  \bold{In this quadrant,
44 of the 86 segments are isothermal.}
The statistics from this graph are
summarized in Table~\ref{the_table}.  
The isothermal segments are among the most massive filaments we
have identified and thus would be expected to be avenues of
transport for a disproportionately large fraction of the gas.
With this in mind, Table~\ref{the_table} shows the fractions by 
gas mass in parentheses.

Despite our efforts to qualify segments, isothermal
cylinders might still fail to \bold{be identified} for a variety of
reasons. In additon to misalignment and extraneous structures
nearby, some of the segments might have a scale radius too small
or too large to be modeled in our study.  With these considerations
in mind, we conclude that isothermal cylinders are at least common among
filament segments residing in regions conducive to the formation
of these structures.

Figure~\ref{shell_final} shows that among the densest shells,
those with an overdensity greater than twenty, about 40\% have
at least one isothermal segment.  

\bold{
The mass per unit length of an isothermal cylinder is finite 
and completely determined by the temperature and ionization
state of the gas \citep{stod63,ostriker64}. 
Excess gas is expected to disrupt an isothermal
cylinder \citep{inutsuka92}.
The isothermal segments
in our simulation are among the most massive and gas-rich of the
simulation  The segments that fail to be isothermal generally have
too little gas to be isothermal, as shown in Figure~\ref{ten}.
Thus in our simulation, at least, the amount of gas per unit length
appears to be limited to the amount compatible with an isothermal
cylinder.  \footnote{\bold{Although the total mass per unit length is
finite, the radial profile extends to infinity.
The profile will thus be truncated at some radial distance,
perhaps by shocks or ram pressure, thus
limiting the linear density  of gas in an isothermal structure 
to a lesser value.
Our potential profile fitting extends to 4.25 scale radii, a
distance enclosing 69\% of the total mass.
The correction is thus modest.}}
However, the high resolution and consequently modest box size
($8h^{-1} \unit{Mpc}$ comoving)
of our simulation limits the mass of galaxies that can be
modeled.  Filaments associated with larger galaxies, particularly
those surrounded by hot halo gas  \citep{dekel09,
keres09,faucher11},
may have different structures from those in our study.
Our findings do not apply to the spherically
symmetric ``hot mode'' accretion that becomes increasingly important
for larger galaxies at lower redshifts
\citep{birnboim03,katz03,keres05,dekel06,
ocvirk08,keres09b,dekel09d,brooks09}. }

From the isothermal potential profile we can determine the central
gas density, a useful value that is difficult to compute
directly from the gas particles of the simulation.  
Figure~\ref{rn} shows that the isothermal
segments have central densities
equal to about 500 times the mean overall cosmic 
density, above the value of 200 normally considered
indicative of \bold{spherically symmetric} virialization \bold{and
well above the corresponding value of about 80 for cylindrical
symmetry \citep{fillmore84}.} \bold{For comparison, the average density 
of gas enclosed within a radius that contains
half the total gas is reduced from the central one by only a factor
of four.}  Many of the segments to the left of
the green line, denoting an overdensity of 200, are isothermal-like 
segments with too little gas to be isothermal. 

\bold{This result suggests that dark matter 
is not needed to maintain the isosthermal 
structure in the presence of the
Hubble expansion.  This finding, coupled with
the high gas fractions,
argues that the structure of these segments is primarily
determined by the hydrodynamic properties of the gas rather 
than by the dark matter.}

\cmdshellfinal
\cmdten

\cmdrn
\cmdvelz

\bold{The filaments appear to extend nearly to the centers of the
galaxies.  On average the gas is falling into the galaxies along
the filaments (Figure~\ref{velz}). However, the 
velocity dispersion is
large, and there is no clear relation to the masses of the 
galaxies.  The nature of the simulation does not allow one
to trace the history of individual gas particles,
and the peculiar velocities of the particles
include that of significant bulk 
motion of larger regions, which must be subtracted out.
Consequently, a more detailed analysis of gas transport 
has not been pursued.  Accretion of filament gas onto galaxies
has been shown directly by
others however \citep{keres05,brooks09,dekel09d}.}

\section{Methods}
\label{methods}

\subsection{Review of Simulation}
\label{simulation}
The simulation from which the filaments are drawn is the same as that in Paper I,
and has been previously described there.
It utilizes a ``Softened
Lagrangian Hydrodynamics'' (SLH-P$^{3}$M) code 
\citep{gnedin95,gnedin_bertschinger_96} with a flat $\Lambda$CDM cosmology.
The cosmological parameters are  $\Omega_{m} = 0.27$, $\Omega_{b} = 0.04$, 
$\sigma_{8} = 0.91$, and $h = 0.71$. 
The simulation is developed following the gas
dynamics on a quasi-Lagrangian mesh, which deforms adaptively to
provide higher resolution in higher density regions.  As in Paper I we 
concentrate on a snapshot of the simulation at a redshift of 5.135.
The softening length was 0.08 proper kpc at this redshift.
\bold{Each dimension of the simulation box is $8h^{-1} \unit{Mpc}$ comoving
and contains 256 cells.}
The dark matter particle mass is $2.73\times10^{6}\dim{M}_\odot$,
and the fiducial mass of a gas particle is
$4.75\times 10^{5}\dim{M}_\odot$.  \bold{Gas particle masses are adjusted
during the simulation by a factor of a few.} 
\bold{Galaxies were identified with DENMAX
\citep{bertschinger91}.}

\subsection{Finding Filaments}
\label{find}
To implement our filament definition \bold{described at the beginning
of section~\ref{results}}
we used the HealPix\footnote{We gratefully acknowledge the use of 
the HealPix software
package obtained from http://healpix.jpl.nasa.gov.} partition 
of the volume of a sphere \citep{gorski05} 
centered on each of 200 galaxies.
A subdivison \bold{(nside = 8)} into 768 \bold{volume elements, called pixels,} proved ideal.

\cmdrangefract

Each angular pixel is divided into four radial
subpixels corresponding to four spherical shells bracketed by 
five spheres at \bold{radii of} 3, 4, 5, 6, and
7 radial units (7.24 kpc) from the center of the galaxy.
\bold{The spherical shell 
in which a segment resides will be termed its parent shell.}

For our study we selected as possible
segments of
filaments those subpixels having at least 30 times the average 
amount of gas per subpixel in the parent shell.
Neighboring subpixels are consolidated
into a single peak segment. 

We then require that the gas be uniformly distributed along
the axis of the segment in order to eliminate galaxies,
grossly misaligned filaments, and other miscellaneous oddities.
For this test, the segment is 
divided longitudinally into three parts,
and the total gas mass within four kpc of the axis is computed for each.
A range fraction is obtained by dividing the maximum
difference between any two parts by the average of the three parts.
We consider the distribution uniform if the range fraction does
not exceed 0.4.  

The 618 segments that meet both tests constitute
the candidate segments to be examined for isothermal structure.
They \bold{extend} from
178 of the 200 galaxies analyzed.

The method, though crude, quickly identifies 
filamentous structures.
Visual inspection shows the segments lined up into
filaments of up to four segments.  On a larger scale the filaments
merge to connect many galaxies in a filamentary network.

A corresponding range fraction was computed for the dark matter in
these candidate segments.  Figure~\ref{range_fract} shows that the gas and
dark matter are distributed very differently by this
criterion.

\bold{
Our method for identifying filaments has the
disadvantage that it requires that the filament issue
from the center of the galaxy in a straight line.
However, visual inspection of the simulation 
suggests that
most of the filaments near galaxies do meet this
requirement.  Our method has the advantage of being
easy to implement on the small scale of our
structures.}


\bold{Several more general methods to find filaments 
have been developed by others for
the analysis of the large scale structure of the 
universe.  Methods based upon Minkowski functionals
\citep{sahni98,schmalzing99}
are promising for our purposes although they are
more difficult to implement.
Others based upon the analysis of first and second
derivatives of the density 
\citep{sousbie08,bond10}
or the analysis of orbits of
test particles \citep{hahn07a}
are not practical for
the small scale of our structures.}  


\subsection{Computing the Gas Potential}
\label{potential}
The gravitational potential generated by the gas of a segment 
is computed on a 
200 x 200 plane grid with a
spacing of 0.17 kpc, perpendicular to the segment at its center.  
The potential is a sum over the contributions of all gas particles 
within 7 kpc of the cylindrical axis.  
Using a preliminary computation, 
this axis is adjusted
to the common center of the largest contours, which are nearly 
concentric circles.

\cmdcontours

The 7 kpc maximal radius for the gas was chosen from experience,
coupled with our findings in Paper I.  A softening
length of 0.1 kpc is added in quadrature to each
contribution.  In order to compare the potential of the finite segment
to the simple form for an infinite cylinder,
the gas particle distribution is replicated ten times in
each direction along the axis, and the potential contributions of these
additional particles added in.

Fifty equally spaced contours on the full grid are computed
using the IDL Software
(Research Systems).
Example potential contours are shown in
Figure~\ref{contours}.

\subsection{Fitting to an Isothermal Potential}
\label{fitting}
The result is a set of contours for a nearly cylindrically symmetric 
potential well.  The average radius of each contour about its own center
is used to construct a radial potential profile.  As will be shown, the
contour centers very nearly coincide.

The zero point for the potential of an isothermal cylinder is defined to 
be the potential on the cylinder axis \citep{ostriker64}.
We take as our estimate of this value the value at the most
extreme grid point within the innermost contour.

An isothermal potential profile is completely characterized by
two parameters, the scale radius and the sound speed.
We determined the scale radius for which the profile is
the best fit up to a constant of proportionality.
The constant for the best fit is then the
square of an ``effective'' sound speed 
\bold{(see Section~\ref{ration}).}

The test region for each trial is a cylindrical shell between one and
4.25 scale radii.    This region
contains the characteristic inflection point of the potential
curve at the square root of eight times the scale radius, a
radius enclosing half the total mass of an isothermal cylinder.
The central region where
small irregularites have a large effect on the profile is thus omitted
from the fitting procedure.

For each of a range of scale radii the standard deviation of the
mean of the proportionality constant throughout the fitting region 
is determined.  The scale radius for which the ratio of this 
standard deviation to the
constant of proportionality itself is a minimum is taken as the
best fitting scale radius.  We require the minimum ratio to be less than 0.01
because we find empirically that these fits narrowly prescribe 
the optimal scale radius. Figure~\ref{fit}, center panel, shows an
example of the determination of the optimal scale radius.
Fits this good usually show the inflection point well.

Segments that fit this well are then further tested by computing
the fractional standard deviation of the fit to the derivative.  
Here, a cutoff of 0.1 was used.  This test eliminates some 
pathological cases, often
involving very small scale radii and thus very small fitting regions.  
Examples of well-fitting profiles of the potential
and its derivative are shown in Figure~\ref{fit}, left and right
panels respectively.

As a further test of the fit, the total mass within four scale radii 
was compared to
prediction.  The computed mass is invariably smaller than predicted
presumably because the centering varies slightly along the axis
of the cylinder.
Only when all centers coincide would the mass be expected to be
100\% that of prediction.  We require that the actual mass be at least
70\% of prediction.  The segments failing this test generally have
very small scale radii and thus a small spatial region for the fitting.

\cmdfit
\cmdquality

Those segments meeting all of these requirements are considered to be
isothermal-like, that is, they have an isothermal profile for
some effective temperature. The number of equally spaced contours
in the fitting region for the best fit ranges from about 20 to 35.
Figure~\ref{quality}, left panel, shows a histogram of their 
scale radii.  Most are between 0.8 and 1.5 kpc.   The low end
may be underrepresented because of the small volume of the
resultant fitting region.  The high end may be underrepresented because
the fitting region is slightly truncated in those few cases where it
would otherwise extend beyond 7 kpc.

\bold{The resolution of the simulation is limited by the discrete number
of gas particles.  It ranges from about 0.5 kpc at the center of the
filament to about 1 kpc at the furthest edge of the fitting region.
The total number of gas particles used to fit a segment profile
ranged from about a hundred to a thousand.}

For cylindrical symmetry,
the centers of the closed contours should coincide.
For the fitted region of each segment the
standard deviation of the contour centers about the mean of these centers was
computed.  Figure~\ref{quality}, center panel, shows a histogram 
of these standard
deviations for all of the segments fitting the isothermal curve and its
derivative for some effective temperature.  Also for straight cylinders 
the contours should be nearly
circular.  Figure~\ref{quality}, right panel, shows
a histogram of the fractional standard deviations of the mean radius
for these contours.

\subsection{Eliminating Overlaps}
\label{eval}

The fit allows one to compute for each segment an 
effective sound speed.  This is the sound speed that
an isothermal cylinder with this spatial distribution of
gas would have.  This sound speed is then compared to the
actual one in order to identify the isothermal segments.
We designate as isothermal those segments with ratios 
within a factor of two of 1.0.
Out of a total of 618 candidates 87 are isothermal.

To make a more accurate assessment we eliminate overlaps.
Overlapping can occur because
a given gas particle can be part of a segment
issuing from more than one galaxy.  Because of possible
misalignments with the radial directions from the centers of
the galaxies in question, the quality of the 
fitting of a region may differ depending upon which galaxy it is considered
to have emanated from.
For each segment, then, we determine what fraction of the gas 
particles are not held in common with other segments with
better fits.  For this procedure, the measure of fit was taken as 
the absolute value of the deviation from one of the 
ratio of the square of the
actual and effective sound speeds.

The segments, weighted by their non-overlapping fractions,
are added together
to form the histograms \bold{in Figures~\ref{gf_orig}, 
~\ref{tmptr}, ~\ref{hone}, ~\ref{rn},
~\ref{ssq_rat},
and Table~\ref{the_table}.}
In the scatter plots  
only segments whose non-overlapping fractions exceed
one half are included.  

After the elimination of overlaps, the 618 candidates are reduced
to the equivalent of 540 non-overlapping segments,
of which 117 are isothermal-like.  Among the isothermal-like
the equivalent of 80 are isothermal.

\section{Conclusions and Discussion}
\label{discussion}
\bold{From a comological hydrodynamic simulation at 
redshift 5, we find 
that a plausible model for intergalactic filaments
is an isothermal gas cylinder whose structure and
stability are determined primarily by the gravitational and
hydrodynamic properties
of the gas.  The cylinders have a central gas density of several
hundred times the mean total cosmic density, with a peak at about 500.
The average temperature of the gas in the cylinders is 
1-2 times $10^{4}\dim{K}$.  The neutral hydrogen fraction is
generally between 0.01 and 0.02.}

\bold{The box size of $8h^{-1} \unit{Mpc}$ comoving limits the
findings to \bolder{haloes} less than 
$10^{11}\dim{M}_\odot$ in total mass (gas, dark matter,
and stars).}

Our findings fit well into the emerging picture of gas transport into
galaxies.  Except for \bolder{haloes} larger than a few 
times $10^{11}\dim{M}_\odot$ 
at low redshifts, gas is believed to enter galaxies primarily
through intergalactic filaments at temperatures well below the
virial temperature of the galaxy and to never be shock-heated
\citep{birnboim03,katz03,keres05,dekel06,ocvirk08,
keres09b,dekel09d,brooks09}.
The temperatures of our filaments are below a few times
$10^{4}\dim{K}$, mostly well below the estimated virial temperatures of
the galaxies, which range from about $4\times10^{5}\dim{K}$ down to
about $3\times10^{4}\dim{K}$\footnote{The virial temperatures were
computed using a spherically symmetric mass profile within the virial
radius of the galaxy.}.

\bold{The intergalactic medium is thought to
be filamentous, and better knowledge of this 
texture may help to understand
the spectra of high
redshift quasars.  The Lyman alpha forest is the
main observational probe of this tenuous
medium.} 

Many of the galaxies in the simulation, particularly small ones, do
not have recognizable intergalactic filaments attached to them.
For these galaxies the diffuse, hot, ionized gas around them may have
few mechanisms to enter.  Thus the presence or absence of filaments
may effectively divide galaxies into two categories, those that can
efficiently accrete gas and form stars and those that cannot.  If the
baryon-rich cores require a minimum amount of gas for the stability of
an isothermal cylinder, it may be that 
the stellar content of a galaxy is a good
indication for overall cosmic density.
The importance of the gas environment for accretion has been
emphasized by \citet{keres05}.

The dearth of satellite galaxies around the Milky Way is sometimes
cited as a problem for the currently favored $\Lambda$CDM cosmology,
which would predict many more satellite dark haloes.  An absence of
filaments at an earlier stage in cosmic history may have
prevented these haloes from becoming luminous today.  Whatever
the explanation, it is worth noting that the simulation predicts
a clear separation between gas and dark matter at the spatial
dimensions of small galaxies.  We would argue, then, that an
assessment of the number of galaxies that form stars requires
a simulation that at a minimum includes some gas hydrodynamics.

We invariably see our filaments embedded within thin sheets of
gas.  This observation suggests that the filaments might emerge from
the sheets.  \citet{schmidburgk67} has shown that an infinite
isothermal cylinder is but one extreme of a series of 
isothermally balanced structures that range from the infinite
isothermal sheet, through intermediate structures containing
regularly spaced, parallel, embedded filaments with elliptical
cross-sections, to the other extreme of isolated cylinders.

A hydrodynamic origin of the filaments may help to explain 
a general difference between the collapse
of dark matter and gas in our simulation.  The dark matter
can collapse only so far into filaments before fragmenting and
collapsing further into spheroidal structures.  The gas on the
other hand can collapse into denser filaments which are stable to
small perturbations.

\section*{Acknowledgments}
We are grateful to Nickolay Y. Gnedin for providing
the output from the simulation, his visualization software IFRIT,
and helpful comments on
the final manuscript.

\bibliographystyle{mn2e}

\bibliography{harford_v3}

\begin{thebibliography}{}

\bibitem[\protect\citeauthoryear{{Bertschinger} \& {Gelb}}{{Bertschinger} \&
  {Gelb}}{1991}]{bertschinger91}
{Bertschinger} E.,  {Gelb} J.~M.,  1991, Computers in Physics, 5, 164

\bibitem[\protect\citeauthoryear{{Binney}}{{Binney}}{2004}]{binney04}
{Binney} J.,  2004, \mnras, 347, 1093

\bibitem[\protect\citeauthoryear{{Birnboim} \& {Dekel}}{{Birnboim} \&
  {Dekel}}{2003}]{birnboim03}
{Birnboim} Y.,  {Dekel} A.,  2003, \mnras, 345, 349

\bibitem[\protect\citeauthoryear{{Bond}, {Strauss} \& {Cen}}{{Bond}
  et~al.}{2010}]{bond10}
{Bond} N.~A.,  {Strauss} M.~A.,    {Cen} R.,  2010, \mnras, 409, 156

\bibitem[\protect\citeauthoryear{{Brooks}, {Governato}, {Quinn}, {Brook} \&
  {Wadsley}}{{Brooks} et~al.}{2009}]{brooks09}
{Brooks} A.~M.,  {Governato} F.,  {Quinn} T.,  {Brook} C.~B.,    {Wadsley} J.,
  2009, \apj, 694, 396

\bibitem[\protect\citeauthoryear{{Chandrasekhar} \& {Fermi}}{{Chandrasekhar} \&
  {Fermi}}{1953}]{chandra53}
{Chandrasekhar} S.,  {Fermi} E.,  1953, \apj, 118, 116

\bibitem[\protect\citeauthoryear{{Dekel} \& {Birnboim}}{{Dekel} \&
  {Birnboim}}{2006}]{dekel06}
{Dekel} A.,  {Birnboim} Y.,  2006, \mnras, 368, 2

\bibitem[\protect\citeauthoryear{{Dekel}, {Birnboim}, {Engel}, {Freundlich},
  {Goerdt}, {Mumcuoglu}, {Neistein}, {Pichon}, {Teyssier} \& {Zinger}}{{Dekel}
  et~al.}{2009}]{dekel09d}
{Dekel} A.,  {Birnboim} Y.,  {Engel} G.,  {Freundlich} J.,  {Goerdt} T.,
  {Mumcuoglu} M.,  {Neistein} E.,  {Pichon} C.,  {Teyssier} R.,    {Zinger} E.,
   2009, \nat, 457, 451

\bibitem[\protect\citeauthoryear{{Dekel}, {Sari} \& {Ceverino}}{{Dekel}
  et~al.}{2009}]{dekel09}
{Dekel} A.,  {Sari} R.,    {Ceverino} D.,  2009, \apj, 703, 785

\bibitem[\protect\citeauthoryear{{Faucher-Giguere}, {Keres} \&
  {Ma}}{{Faucher-Giguere} et~al.}{2011}]{faucher11}
{Faucher-Giguere} C.,  {Keres} D.,    {Ma} C.,  2011, ArXiv e-prints

\bibitem[\protect\citeauthoryear{{Fillmore} \& {Goldreich}}{{Fillmore} \&
  {Goldreich}}{1984}]{fillmore84}
{Fillmore} J.~A.,  {Goldreich} P.,  1984, \apj, 281, 1

\bibitem[\protect\citeauthoryear{{Gnedin}}{{Gnedin}}{1995}]{gnedin95}
{Gnedin} N.~Y.,  1995, \apjs, 97, 231

\bibitem[\protect\citeauthoryear{{Gnedin}, {Baker}, {Bethell}, {Drosback},
  {Harford}, {Hicks}, {Jensen}, {Keeney}, {Kelso}, {Neyrinck}, {Pollack} \&
  {van Vliet}}{{Gnedin} et~al.}{2003}]{gnedin_filt03}
{Gnedin} N.~Y.,  {Baker} E.~J.,  {Bethell} T.~J.,  {Drosback} M.~M.,  {Harford}
  A.~G.,  {Hicks} A.~K.,  {Jensen} A.~G.,  {Keeney} B.~A.,  {Kelso} C.~M.,
  {Neyrinck} M.~C.,  {Pollack} S.~E.,    {van Vliet} T.~P.,  2003, \apj, 583,
  525

\bibitem[\protect\citeauthoryear{{Gnedin} \& {Bertschinger}}{{Gnedin} \&
  {Bertschinger}}{1996}]{gnedin_bertschinger_96}
{Gnedin} N.~Y.,  {Bertschinger} E.,  1996, \apj, 470, 115

\bibitem[\protect\citeauthoryear{{Gnedin} \& {Fan}}{{Gnedin} \&
  {Fan}}{2006}]{gnedin_fan06}
{Gnedin} N.~Y.,  {Fan} X.,  2006, \apj, 648, 1

\bibitem[\protect\citeauthoryear{{G{\'o}rski}, {Hivon}, {Banday}, {Wandelt},
  {Hansen}, {Reinecke} \& {Bartelmann}}{{G{\'o}rski} et~al.}{2005}]{gorski05}
{G{\'o}rski} K.~M.,  {Hivon} E.,  {Banday} A.~J.,  {Wandelt} B.~D.,  {Hansen}
  F.~K.,  {Reinecke} M.,    {Bartelmann} M.,  2005, \apj, 622, 759

\bibitem[\protect\citeauthoryear{{Hahn}, {Porciani}, {Carollo} \&
  {Dekel}}{{Hahn} et~al.}{2007}]{hahn07a}
{Hahn} O.,  {Porciani} C.,  {Carollo} C.~M.,    {Dekel} A.,  2007, \mnras, 375,
  489

\bibitem[\protect\citeauthoryear{{Harford} \& {Gnedin}}{{Harford} \&
  {Gnedin}}{2003}]{harford03}
{Harford} A.~G.,  {Gnedin} N.~Y.,  2003, \apj, 597, 74

\bibitem[\protect\citeauthoryear{{Harford} \& {Gnedin}}{{Harford} \&
  {Gnedin}}{2007}]{harford07}
{Harford} A.~G.,  {Gnedin} N.~Y.,  2007, \apj, 664, 599

\bibitem[\protect\citeauthoryear{{Harford}, {Hamilton} \& {Gnedin}}{{Harford}
  et~al.}{2008}]{harford08}
{Harford} A.~G.,  {Hamilton} A.~J.~S.,    {Gnedin} N.~Y.,  2008, \mnras, 389,
  880

\bibitem[\protect\citeauthoryear{{Inutsuka} \& {Miyama}}{{Inutsuka} \&
  {Miyama}}{1992}]{inutsuka92}
{Inutsuka} S.,  {Miyama} S.~M.,  1992, \apj, 388, 392

\bibitem[\protect\citeauthoryear{{Katz}, {Keres}, {Dave} \& {Weinberg}}{{Katz}
  et~al.}{2003}]{katz03}
{Katz} N.,  {Keres} D.,  {Dave} R.,    {Weinberg} D.~H.,  2003, in
  {J.~L.~Rosenberg \& M.~E.~Putman} ed., The IGM/Galaxy Connection. The
  Distribution of Baryons at z=0 Vol.~281 of Astrophysics and Space Science
  Library, {How Do Galaxies Get Their Gas}.
pp 185--191

\bibitem[\protect\citeauthoryear{{Kere{\v s}} \& {Hernquist}}{{Kere{\v s}} \&
  {Hernquist}}{2009}]{keres09}
{Kere{\v s}} D.,  {Hernquist} L.,  2009, \apjl, 700, L1

\bibitem[\protect\citeauthoryear{{Kere{\v s}}, {Katz}, {Fardal}, {Dav{\'e}} \&
  {Weinberg}}{{Kere{\v s}} et~al.}{2009}]{keres09b}
{Kere{\v s}} D.,  {Katz} N.,  {Fardal} M.,  {Dav{\'e}} R.,    {Weinberg} D.~H.,
   2009, \mnras, 395, 160

\bibitem[\protect\citeauthoryear{{Kere{\v s}}, {Katz}, {Weinberg} \&
  {Dav{\'e}}}{{Kere{\v s}} et~al.}{2005}]{keres05}
{Kere{\v s}} D.,  {Katz} N.,  {Weinberg} D.~H.,    {Dav{\'e}} R.,  2005,
  \mnras, 363, 2

\bibitem[\protect\citeauthoryear{{Ocvirk}, {Pichon} \& {Teyssier}}{{Ocvirk}
  et~al.}{2008}]{ocvirk08}
{Ocvirk} P.,  {Pichon} C.,    {Teyssier} R.,  2008, \mnras, 390, 1326

\bibitem[\protect\citeauthoryear{{Ostriker}}{{Ostriker}}{1964}]{ostriker64}
{Ostriker} J.,  1964, \apj, 140, 1056

\bibitem[\protect\citeauthoryear{{Sahni}, {Sathyaprakash} \&
  {Shandarin}}{{Sahni} et~al.}{1998}]{sahni98}
{Sahni} V.,  {Sathyaprakash} B.~S.,    {Shandarin} S.~F.,  1998, \apjl, 495, L5

\bibitem[\protect\citeauthoryear{{Schmalzing}, {Buchert}, {Melott}, {Sahni},
  {Sathyaprakash} \& {Shandarin}}{{Schmalzing} et~al.}{1999}]{schmalzing99}
{Schmalzing} J.,  {Buchert} T.,  {Melott} A.~L.,  {Sahni} V.,  {Sathyaprakash}
  B.~S.,    {Shandarin} S.~F.,  1999, \apj, 526, 568

\bibitem[\protect\citeauthoryear{{Schmid-Burgk}}{{Schmid-Burgk}}{1967}]{schmid%
burgk67}
{Schmid-Burgk} J.,  1967, \apj, 149, 727

\bibitem[\protect\citeauthoryear{{Sousbie}, {Pichon}, {Colombi}, {Novikov} \&
  {Pogosyan}}{{Sousbie} et~al.}{2008}]{sousbie08}
{Sousbie} T.,  {Pichon} C.,  {Colombi} S.,  {Novikov} D.,    {Pogosyan} D.,
  2008, \mnras, 383, 1655

\bibitem[\protect\citeauthoryear{{Stod{\'o}lkiewicz}}{{Stod{\'o}lkiewicz}}{196%
3}]{stod63}
{Stod{\'o}lkiewicz} J.~S.,  1963, Acta Astronomica, 13, 30

\end{thebibliography}

\end{document}